\documentclass[12pt]{article}
\usepackage{amsmath}
\usepackage{amssymb}
\usepackage{bm}% bold math
\usepackage{epsfig, graphicx, euscript}
\newcommand{\be}{\begin{equation}}
\newcommand{\ee}{\end{equation}}
\newcommand{\bea}{\begin{eqnarray}}
\newcommand{\eea}{\end{eqnarray}}

\newcommand{\bfn}{\mbox{\boldmath $n$}}
\newcommand{\bfr}{\mbox{\boldmath $r$}}
\newcommand{\mbss}[1]{_{\mbox{\scriptsize #1}}}
\newcommand{\mbts}[1]{_{\mbox{\tiny #1}}}
\newcommand{\mbsu}[1]{\mbox{\scriptsize #1}}

\newcommand{\ds}{\displaystyle}
\newcommand{\scs}{\scriptstyle}
\newcommand{\txts}{\textstyle}
\newcommand{\ve}{\varepsilon}
\newcommand{\vk}{\varkappa}
\newcommand{\vphi}{\varphi}

\makeatletter
\renewcommand{\section}{\@startsection{section}{1}{0pt}%
{-3.5ex plus -1ex minus -.2ex}{2.3ex plus.2ex}%
{\normalsize\bf}}
\renewcommand{\subsection}{\@startsection{subsection}{1}{0pt}%
{-3.5ex plus -1ex minus -.2ex}{2.3ex plus.2ex}%
{\normalsize\bf}}
\makeatother
\begin{document}
\title{%
\large\bf
DESCRIPTION OF ELECTRIC DIPOLE EXCITATIONS\\
IN THE TIN ISOTOPES WITHIN THE\\
QUASIPARTICLE TIME BLOCKING APPROXIMATION}
\author{%
 E. V. Litvinova$\,^{1,2}\,$ and
 V. I. Tselyaev$\,^3$\\
{\it \normalsize
 $^1\!\!$
 Institute of Physics and Power Engineering,
 249020, Obninsk, Russia}$\vphantom{,}$\\
{\it \normalsize
 $^2\!\!$
 Physik-Department der Technischen Universit\"at M\"unchen,}\\
{\it \normalsize
 D-85748 Garching, Germany}$\vphantom{,}$\\
{\it \normalsize
 $^3\!$
 Nuclear Physics Department,
 V. A. Fock Institute of Physics,}\\
{\it \normalsize
 St. Petersburg State University, 198504,
 St. Petersburg, Russia}}
\date{\today}
\maketitle
\begin{abstract}
The quasiparticle time blocking approximation (QTBA)
is applied to describe E1 excitations in the even-even tin
isotopes. Within the model pairing correlations,
two-quasiparticle (2q), and 2q$\otimes$phonon
configurations are included. Thus the QTBA is
an extension of the quasiparticle random phase approximation
to include quasiparticle-phonon coupling.
Calculational formulas are presented in case of neutral
excitations in the spherically symmetric system.
The main equations are written in the coordinate representation
that allows to take into account single-particle continuum
completely.
The E1 photoabsorption cross sections have been calculated
in nuclei $^{116,120,124}$Sn. It has been obtained that the
2q$\otimes$phonon configurations provide noticeable
fragmentation of the giant dipole resonance resulting in
appearance of significant spreading width. The results are
compared with available experimental data.
\vspace{2em}
\begin{flushleft}
\end{flushleft}
\end{abstract}
\newpage
\section{INTRODUCTION}

Theoretical description of giant multipole resonances (GMRs)
and resonance structures in magic and open-shell nuclei
has a long history and remains a problem of current importance.
The main tools in solving this problem within the framework of
microscopic approach are the random phase approximation (RPA)
and the quasiparticle RPA (QRPA) which is straightforward
generalization of the RPA to include pairing correlations.
However, despite these models enable one to describe
mean energies and total strengths of the GMRs,
they fail to reproduce the total widths of the resonances
and their fine structure (see, e.g., Ref.~\cite{KTT}).
One of the reasons is that RPA and QRPA do not provide a
mechanism producing spreading width $\Gamma^{\downarrow}$
which gives a considerable contribution in the total widths
of the GMRs.

Simplest mechanism of this type is the coupling of the
quasiparticles to phonons being superpositions of the
one-particle-one-hole (1p1h) or the two-quasiparticle (2q)
configurations. As applied to structure of the even-even
nuclei, the concept of the quasiparticle-phonon coupling
(QPC, see Ref.~\cite{BM2}) enables one to take into account
1p1h$\otimes$phonon and 2q$\otimes$phonon configurations
in addition to 1p1h and 2q ones incorporated within the
RPA and the QRPA. A series of models has been developed to
go beyond the RPA and the QRPA by means of inclusion of this
mechanism (see Refs. \cite{KTT}, \cite{S92}--\cite{KL04} and
references therein).
Recently, new model has been formulated (see Ref.~\cite{T05})
in which pairing correlations, 2q, and 2q$\otimes$phonon
configurations are included. This model is based on the Green
function (GF) formalism that mainly determines its name:
quasiparticle time blocking approximation (QTBA).
The QTBA is a generalization of the method of chronological
decoupling of diagrams (MCDD) developed in Ref.~\cite{T89}
to describe excited states of the even-even nuclei
without pairing.

The first aim of this paper is
to present calculational formulas obtained from the
general ones of the model \cite{T05} making use
of certain approximations in case of neutral excitations in
the spherically symmetric system.
In particular, in the formulas presented
zero-range forces are adopted as an effective interaction
and the Bardeen-Cooper-Schrieffer (BCS) approximation is used
to determine quasiparticle energies and wave functions.
The equations obtained are a system of coupled equations
corresponding to the excitations in particle-hole,
particle-particle, and hole-hole channels.
The basic equations are written in the coordinate representation
that allows to take into account single-particle continuum
completely. Notice that inclusion of the continuum is of particular
importance to describe correctly the widths of the GMRs.

Our second aim is to test the QTBA in calculations
of the electric dipole excitations in the nuclei with pairing.
With that end in view we have chosen tin isotopes
$^{116,120,124}$Sn.
We present E1 photoabsorption cross sections calculated within
QRPA and QTBA. Calculated integral characteristics of the
giant dipole resonance (GDR)
are compared with the experimental data.

The paper is organized as follows.
In Sec.~\ref{sect2} general scheme of the approach is presented.
In Sec.~\ref{sect3} the basic equations of the QTBA
are transformed to channel form in the coordinate representation.
In Sec.~\ref{sect4} these equations are further reduced
to the equations for partial components.
Formulas for the correlated propagator of the QTBA
in the coordinate representation are presented in Sec.~\ref{cpcr}.
In Sec.~\ref{qpi} the equations for the matrix elements
of the amplitude of quasiparticle-phonon interaction
are obtained. These quantities enter the equations
for the correlated propagator of the QTBA in terms
of the reduced matrix elements which are presented
in Sec.~\ref{proprme}.
In Sec.~\ref{results} we describe our calculational scheme and
present the results obtained for the photoabsorption cross sections
in three tin isotopes.
Conclusions are drawn in the last section.

\section{GENERAL FRAMEWORK \label{sect2}}

The basic quantity, which determines the physical
observables in the QTBA, is the nuclear polarizability
$\Pi (\omega)$. More precisely, it determines distribution of
the transition strength caused by an external field
$V^{\,0}$. In the representation of the
single-quasiparticle basis functions, which
will be specified below, $\Pi (\omega)$ is defined as
\be
\Pi (\omega) = - \frac{1}{2}\,\sum_{1234}\,
(eV^{\,0})^{\ds *}_{21}
\,R^{\,\mbsu{eff}}_{12,34} (\omega)\,
(eV^{\,0})^{\vphantom{\ds *}}_{43}\,,
\label{dfpol1}
\ee
where $e$ is the effective charge operator,
$\omega$ is complex energy variable,
$R^{\,\mbsu{eff}}(\omega)$ is the effective response function.
$R^{\,\mbsu{eff}}(\omega)$ is a solution of the following
Bethe-Salpeter equation
\be
R^{\,\mbsu{eff}}_{12,34} (\omega) =
A^{\vphantom{\mbsu{eff}}}_{12,34} (\omega) -
\sum_{5678} A^{\vphantom{\mbsu{eff}}}_{12,56} (\omega)\,
{\cal F}^{\vphantom{\mbsu{eff}}}_{56,78}\,
R^{\,\mbsu{eff}}_{78,34} (\omega)\,,
\label{bseqren}
\ee
where $A(\omega)$ is a correlated propagator, ${\cal F}$
is an amplitude of the effective interaction.
In particular, the strength function $S(E)$ which is frequently
used for the description of nuclear excitations
is expressed in terms of the polarizability as
\be
S(E) = - \frac{1}{\pi}\,\mbox{Im}\,\Pi(E + i\,\Delta)\,,
\label{dfstrf}
\ee
where $\Delta$ is a smearing parameter.

The Eq.~(\ref{bseqren}) for the effective response function
is quite general. It was obtained in Ref.~\cite{T05} for
the Fermi systems with pairing within the framework of
generalized Green function formalism in which the normal and
the anomalous GFs are treated in a unified way in terms
of the components of generalized GFs in a doubled space.
The physical content of the model
is determined by the choice of the propagator $A(\omega)$.
In particular, the QRPA equation for the $R^{\,\mbsu{eff}}(\omega)$
has the same form (\ref{bseqren}) with uncorrelated
propagator taken instead of $A(\omega)$.
The equations defining $A(\omega)$ within the QTBA are given
in Ref.~\cite{T05}. In the present paper we consider
the version of QTBA in which the correlated propagator includes
contributions of the 2q and 2q$\otimes$phonon configurations.

As a first step we have to determine the single-quasiparticle
basis functions $\psi_1(y)$ which form matrix representation
of the theory. For the Fermi systems with pairing
these functions are defined in the doubled space
spanned by the coordinates $y = \{x, \chi \}$,
where symbol $x = \{\bfr, \sigma, \tau \}$ includes the
spatial coordinate $\bfr$, the spin $\sigma$, and the isospin
$\tau$ variables, $\chi = \pm 1$ is an additional index
introduced for denoting the different components of the
single-quasiparticle functions in the doubled coordinate space.
Index $1 = \{\lambda_1, \eta_1 \}$ of the doubled configuration
space includes index $\lambda_1$ of the usual single-particle
configuration space and the index $\eta_1 = \pm 1$ which is
the sign of the eigenvalue corresponding to $\psi_1(y)$.
Namely, we have: ${\cal H}\,\psi_1 = E_1\,\psi_1$ where
${\cal H}$ is single-quasiparticle Hamiltonian,
$E_1 = \eta_1\,E_{\lambda_{\mbts{1}}}$,
$\,E_{\lambda_{\mbts{1}}} = |E_1|$.
In case of the spherically symmetric system,
we are interested in, the index
$\lambda$ can be represented by the following set:
$\lambda = \{(\lambda), m \}$
where $(\lambda) = \{\tau_{\lambda}, n, l, j \}$, $m$ is
the projection of the total angular momentum,
and we have: $E_{\lambda} = E_{(\lambda)}$.

We will use the approximation corresponding
to so-called canonical basis representation of the functions
$\psi_1(y)$ (see Ref.~\cite{KSTV} for details).
To determine functions $\psi_1(y)$ within this approximation
let us note that in the matrix form the Hamiltonian ${\cal H}$
reads
\be
{\cal H} =
\left(
\begin{array}{cc}
h - \mu & \Delta \\ - \Delta^* & \mu -h^* \\
\end{array}
\right)
\label{eqm}
\ee
where $h = h (x,x')$ is the single-particle Hamiltonian,
$\Delta = \Delta (x,x')$ is the operator of the pairing field,
$\mu$ is the chemical potential.
Let $\{\vphi^{\vphantom{*}}_{\lambda}(x)\}$ be the complete set
of orthonormal eigenfunctions of the Hamiltonian $h (x,x')$:
$\;h\,\vphi^{\vphantom{*}}_{\lambda} =
\ve^{\vphantom{*}}_{\lambda}\,\vphi^{\vphantom{*}}_{\lambda}$.
We will assume that the operator $\Delta (x,x')$ has the
canonical form in the same basis
$\{\vphi^{\vphantom{*}}_{\lambda}(x)\}$ that corresponds to
the state-dependent version of the BCS approximation
(see, e.g., Ref.~\cite{RS}), or to
the so-called approximation of the diagonal pairing.
In this case for the spherically symmetric system we have:
\be
\left.
\begin{array}{rcl}
\psi_{\lambda,+}(x,+) = u^{\vphantom{*}}_{\lambda}\,
\vphi^{\vphantom{*}}_{\lambda}(x)\,,\quad &
\psi_{\lambda,+}(x,-) = (-1)^{l+j+m}\,
v^{\vphantom{*}}_{\lambda}\, \vphi^*_{\bar{\lambda}}(x)\,,\\
\psi_{\lambda,-}(x,-) = u^{\vphantom{*}}_{\lambda}\,
\vphi^*_{\lambda}(x)\,,\quad &
\psi_{\lambda,-}(x,+) = (-1)^{l+j+m}\,
v^{\vphantom{*}}_{\lambda}\,
\vphi^{\vphantom{*}}_{\bar{\lambda}}(x)\,,\\
\end{array}
\right\}
\label{uvpsisa}
\ee
where $\bar{\lambda} = \{(\lambda), -m \}$,
$v_{\lambda}$ and $u_{\lambda}$ are real
numbers which satisfy the following conditions:
$\,u^{\vphantom{2}}_{\lambda} =
\sqrt{1 - v^2_{\lambda} \vphantom{V^A}} \geqslant 0\,$,
$\,v^2_{\vphantom{\bar{\lambda}}\lambda} =
v^2_{\bar{\lambda}} \leqslant 1$
[see Eqs. (\ref{uvfor})--(\ref{fxill}) for an explicit
definition of $v_{\lambda}$ and $u_{\lambda}$ within
the BCS approximation].
The choice of the phase
factors is determined by the formulas:
\bea
\vphi^{\vphantom{*}}_{\lambda}(x) &=&
\delta_{\tau_{\lambda},\tau}\, R_{(\lambda)} (r)\,
\phi^{\vphantom{*}}_{jlm} (\bfn, \sigma)\,,
\label{dfphil}\\
\phi^{\vphantom{*}}_{jlm} (\bfn, \sigma) &=&
\sum_{\mu} (l\mu\, {\txts \frac{1}{2}}\sigma | j m)\,
Y_{l\mu} (\bfn)\,,
\label{dfsaf}\\
\vphi^{\vphantom{*}}_{\lambda}(x) &=&
(-1)^{l+j+m+\frac{1}{2}+\sigma}\,
\vphi^*_{\bar{\lambda}}(\bar{x})\,,
\label{relphi}
\eea
where $\bfn = \bfr / r$, $\,\bar{x} = \{\bfr, -\sigma, \tau\}$.

In applications of the theory it is convenient to use
another basis functions which differ from the functions
(\ref{uvpsisa}) by a unitary transformation. Let us
introduce a matrix
\be
{\cal O}_{12} =
{\cal O}_{\lambda_{\mbts{1}}\eta_{\mbts{1}},\,
\lambda_{\mbts{2}}\eta_{\mbts{2}}} =
\delta_{\eta_{\mbts{1}},\,\eta_{\mbts{2}}}\,
[\,\delta_{\eta_{\mbts{1}},\,+1}\,
\delta_{\lambda_{\mbts{1}},\,\lambda_{\mbts{2}}} +
(-1)^{l_{\mbts{1}}+j_{\mbts{1}}-m_{\mbts{1}}}\,
\delta_{\eta_{\mbts{1}},\,-1}\,
\delta_{\lambda_{\mbts{1}},\,\bar{\lambda}_{\mbts{2}}}]\,.
\label{dfmo}
\ee
This matrix is real and orthogonal, and consequently
it is unitary. So wave functions $\tilde{\psi}_1(y)$
defined through the single-quasiparticle basis functions
$\psi_1(y)$ by the formula
\be
\tilde{\psi}_1(y) = \sum_2 {\cal O}_{12}\,\psi_2(y)
\label{dfpsit}
\ee
also form a complete set of the orthonormal functions.
We will use just the set $\{\tilde{\psi}_1(y)\}$
as the set of basis functions.
This does not lead to an inconsistency since
the single-particle GF $\tilde{G}(\ve) = (\ve - {\cal H})^{-1}$
is diagonal both in $\{\psi_1(y)\}$ and in $\{\tilde{\psi}_1(y)\}$
representation, and hence
the formulas for the correlated propagator (see Ref.~\cite{T05})
are the same in both representations.

To describe dynamics of the system and to calculate
the polarizability and the strength function we start from
the equation for linear response matrix (LRM) $\Lambda$.
Notice that in the present work the term LRM
is used instead of the frequently used one
``density matrix variation in an external field''
because it is more correct in our notations.
In the coordinate representation the equation for $\Lambda$
is obtained by the convolution of the equation for effective
response function $R^{\,\mbsu{eff}}$ with an operator of the
renormalized external field $eV^{\,0}$:
\bea
\Lambda (y_1,y_2;\,\omega) &=& \Lambda^0 (y_1,y_2;\,\omega) -
\int dy_3\,dy_4\,dy_5\,dy_6\,A (y_1,y_2;\,y_3,y_4;\,\omega)
\nonumber\\
&\times& {\cal F} (y_3,y_4;\,y_5,y_6)\,
\Lambda (y_5,y_6;\,\omega)\,,
\label{lrmeq}
\eea
where
\be
\Lambda (y_1,y_2;\,\omega) = - \sum_{1234}
\tilde{\psi}^*_1(y_1)\,\tilde{\psi}^{\vphantom{*}}_2(y_2)\,
R^{\,\mbsu{eff}}_{12,34} (\omega)\,
(eV^{\,0})^{\vphantom{*}}_{43}\,,
\label{dflrm}
\ee
\bea
\Lambda^0 (y_1,y_2;\,\omega) &=& - \int dx_3\,dx_4\,
[\,A (y_1,y_2;\,x_3+,x_4+;\,\omega) -
   A (y_1,y_2;\,x_4-,x_3-;\,\omega)\,]
\nonumber\\
&\times& \tilde{V}^0 (x_4,x_3)\,.
\label{dflrm0}
\eea
It is assumed that
the correlated propagator of the model $A(\omega)$ is
initially calculated in configuration space and then is
transformed to coordinate representation:
\be
A (y_1,y_2;\,y_3,y_4;\,\omega) = \sum_{1234}
\tilde{\psi}^*_1(y_1)\,\tilde{\psi}^{\vphantom{*}}_2(y_2)\,
\tilde{\psi}^{\vphantom{*}}_3(y_3)\,
\tilde{\psi}^*_4(y_4)\,A_{12,34} (\omega)\,.
\label{propcr}
\ee
Components of the external field are
\be
\tilde{V}^0 (x_1,x_2) = \tilde{V}^0 (x_1+,x_2+) =
-\tilde{V}^0 (x_2-,x_1-) = \sum_{12}
\tilde{\psi}^{\vphantom{*}}_1(x_1+)\,\tilde{\psi}^*_2(x_2+)\,
(eV^{\,0})^{\vphantom{*}}_{12}\,,
\label{dfexfr}
\ee
and it is supposed that
$\tilde{V}^0 (x_1+,x_2-)=\tilde{V}^0 (x_1-,x_2+)=0$.

In terms of the LRM the Eq.~(\ref{dfpol1}) for the
polarizability reads:
\be
\Pi (\omega) = \int  dx_1\,dx_2\,
\tilde{V}^{0^{\scs *}} (x_2,x_1)\,
\Lambda (x_1+,x_2+;\,\omega)\,.
\label{dfpol3}
\ee

To determine general form of the effective interaction in our
approach let us note that within a self-consistent scheme the
amplitude ${\cal F}$ in Eq.~(\ref{lrmeq}) can be defined as
a second order functional derivative of some energy density
functional ${\cal E}[{\cal R}]$ where ${\cal R}$ is the extended
density matrix
($\,{\cal R}(y_1,y_2) = \sum\limits_1
\delta^{\vphantom{A}}_{\eta_{\mbts{1}},\,-1}\,
\tilde{\psi}^{\vphantom{*}}_1(y_1)\,\tilde{\psi}^*_1(y_2)$,
see, e.g., Ref.~\cite{KSTV}):
\be
{\cal F} (y_1,y_2;\,y_3,y_4) =
\frac{2\,\delta^2 {\cal E}[{\cal R}]}
{\delta {\cal R}(y_1,y_2)\,\delta {\cal R}(y_4,y_3)}\,.
\label{dffsc}
\ee
If ${\cal E}[{\cal R}]$ is usual functional
of the Hartree-Fock-Bogoliubov theory built up
on the basis of the Hamiltonian which includes only two-particle
density-independent interaction with the antisymmetrized
amplitude $w^{(2)}$, we have:
\bea
&&{\cal F} (y_1,y_2;\,y_3,y_4) =
{\txts \frac{1}{2}}\,
\delta_{\chi_{\mbts{1}},\,\chi_{\mbts{2}}}\,
\delta_{\chi_{\mbts{3}},\,\chi_{\mbts{4}}}
\nonumber\\
&&\times \biggl(\,
\delta_{\chi_{\mbts{1}},\,\chi_{\mbts{3}}}\,\bigl[\,
\delta_{\chi_{\mbts{1}},\,+1}\,
{\cal F}^{+} (x_1,x_2;\,x_3,x_4) +
\delta_{\chi_{\mbts{1}},\,-1}\,
{\cal F}^{+} (x_2,x_1;\,x_4,x_3)\,\bigr]
\nonumber\\
&&-\,
\delta_{\chi_{\mbts{1}},\,-\chi_{\mbts{3}}}\,\bigl[\,
\delta_{\chi_{\mbts{1}},\,+1}\,
{\cal F}^{+} (x_1,x_2;\,x_4,x_3) +
\delta_{\chi_{\mbts{1}},\,-1}\,
{\cal F}^{+} (x_2,x_1;\,x_3,x_4)\,\bigr]\,\biggr)
\nonumber\\
&&+\,
\delta_{\chi_{\mbts{1}},\,-\chi_{\mbts{2}}}\,
\delta_{\chi_{\mbts{3}},\,-\chi_{\mbts{4}}}
\delta_{\chi_{\mbts{1}},\,\chi_{\mbts{3}}}\,\bigl[\,
\delta_{\chi_{\mbts{2}},\,+1}\,
{\cal F}^{-} (x_1,x_2;\,x_3,x_4) +
\delta_{\chi_{\mbts{2}},\,-1}\,
{\cal F}^{-} (x_3,x_4;\,x_1,x_2)\,\bigr]\,,
\label{inthfb}
\eea
where
\be
{\cal F}^{+} (x_1,x_2;\,x_3,x_4) =
w^{(2)} (x_2,x_3;\,x_1,x_4)\,,\quad
{\cal F}^{-} (x_1,x_2;\,x_3,x_4) = {\txts \frac{1}{2}}\;
w^{(2)} (x_1,x_2;\,x_3,x_4)\,.
\label{intpm}
\ee
In what follows we assume that Eq.~(\ref{inthfb})
is fulfilled for the interaction ${\cal F}$,
however Eqs.~(\ref{intpm}) are not
supposed to be fulfilled. In other words, the amplitudes
${\cal F}^{+}$ and ${\cal F}^{-}$ will be considered
to be independent, and it will be supposed that no other
independent components are contained in the amplitude
${\cal F}$. These assumptions correspond to the
approximation adopted in the Theory of Finite Fermi Systems
with pairing correlations (TFFSPC, Ref.~\cite{M67}).

\section{TRANSFORMATION OF THE EQUATION\\
         FOR LINEAR RESPONSE MATRIX TO CHANNEL FORM
         \label{sect3}}

The LRM defined by Eq.~(\ref{dflrm}) contains information
about excitations of the initial system in three different
channels corresponding to the transitions to the states of
the final systems with different numbers of particles.
Suppose that the number of particles in the ground state of
the initial system is conserved and is equal to $N_0$.
Let $N$ be the number of particles in the final system.
Then, in accordance with the standard terminology, we have:
(i) ph channel if $N=N_0\,$;
(ii) pp channel if $N=N_0+2\,$;
(iii) hh channel if $N=N_0-2$.
Notice that it is not necessary to introduce
hp channel explicitly because of symmetry of the LRM
and other quantities.

Let us introduce the channel index $c \in \{ ph,\,pp,\,hh\}$
and define the projection operators $\Xi^{(c)}$:
\bea
\Xi^{(ph)} (x_1,x_2;\,y_3,y_4) &=&
\delta_{\chi_{\mbts{3}},\,+1}\,
\delta_{\chi_{\mbts{4}},\,+1}\,
\delta (x_1,\,x_3)\,\delta (x_2,\,x_4)\,,
\label{projph}\\
\Xi^{(pp)} (x_1,x_2;\,y_3,y_4) &=&
\delta_{\chi_{\mbts{3}},\,+1}\,
\delta_{\chi_{\mbts{4}},\,-1}\,
(-1)^{\frac{1}{2}+\sigma_{\mbts{2}}}\,
\delta (x_1,\,x_3)\,\delta (\bar{x}_2,\,x_4)\,,
\label{projpp}\\
\Xi^{(hh)} (x_1,x_2;\,y_3,y_4) &=&
\delta_{\chi_{\mbts{3}},\,-1}\,
\delta_{\chi_{\mbts{4}},\,+1}\,
(-1)^{\frac{1}{2}+\sigma_{\mbts{1}}}\,
\delta (\bar{x}_1,\,x_3)\,\delta (x_2,\,x_4)\,.
\label{projhh}
\eea
The sense of these operators is obvious: acting on any
quasiparticle operator
they cut out its components with fixed $\chi$ values.
Thus applying each of these projectors
to LRM and correlated propagator we obtain
the following components:
\bea
\Lambda^{(c)} (x_1,x_2;\,\omega) &=&
\int dy_3\,dy_4\,\Xi^{(c)} (x_1,x_2;\,y_3,y_4)\,
\Lambda (y_3,y_4;\,\omega)\,,
\label{dflrmc}\\
\Lambda^{0\,(c)} (x_1,x_2;\,\omega) &=&
\int dy_3\,dy_4\,\Xi^{(c)} (x_1,x_2;\,y_3,y_4)\,
\Lambda^0 (y_3,y_4;\,\omega)\,,
\label{dflrm0c}
\eea
\bea
A^{(c,ph)} (x_1,x_2;\,x_3,x_4;\,\omega) &=&
\int dy_5\,dy_6\,\Xi^{(c)} (x_1,x_2;\,y_5,y_6)
\nonumber\\
&\times&
\bigl[\,A (y_5,y_6;\,x_3+,x_4+;\,\omega) -
A (y_5,y_6;\,x_4-,x_3-;\,\omega)\,\bigr]\,,
\label{dfacph}
\eea
\bea
A^{(c,pp)} (x_1,x_2;\,x_3,x_4;\,\omega) &=&
\int dy_5\,dy_6\,\Xi^{(c)} (x_1,x_2;\,y_5,y_6)
\nonumber\\
&\times&
A (y_5,y_6;\,x_3+,\bar{x}_4-;\,\omega)\,
(-1)^{\frac{1}{2}+\sigma_{\mbts{4}}}\,,
\label{dfacpp}\\
A^{(c,hh)} (x_1,x_2;\,x_3,x_4;\,\omega) &=&
\int dy_5\,dy_6\,\Xi^{(c)} (x_1,x_2;\,y_5,y_6)
\nonumber\\
&\times&
A (y_5,y_6;\,\bar{x}_3-,x_4+;\,\omega)\,
(-1)^{\frac{1}{2}+\sigma_{\mbts{3}}}\,,
\label{dfachh}
\eea
where the second channel indices in
the Eqs. (\ref{dfacph})--(\ref{dfachh})
are fixed  by the $\chi$-indices of the propagators
on the right-hand sides of these equations.

Let us also denote:
\bea
{\cal F}^{(c,c')} (x_1,x_2;\,x_3,x_4) &=&
\delta_{c,c'}\,\tilde{{\cal F}}^{(c)} (x_1,x_2;\,x_3,x_4) +
{\cal F}^{\,rest\,(c,c')} (x_1,x_2;\,x_3,x_4)\,,
\label{dffcc}\\
\tilde{{\cal F}}^{(ph)} (x_1,x_2;\,x_3,x_4) &=&
{\cal F}^{+} (x_1,x_2;\,x_3,x_4)\,,
\label{dffcph}\\
\tilde{{\cal F}}^{(pp)} (x_1,x_2;\,x_3,x_4) &=&
(-1)^{\frac{1}{2}+\sigma_{\mbts{2}}+
\frac{1}{2}+\sigma_{\mbts{4}}}\,
{\cal F}^{-} (x_3,\bar{x}_4;\,x_1,\bar{x}_2)\,,
\label{dffcpp}\\
\tilde{{\cal F}}^{(hh)} (x_1,x_2;\,x_3,x_4) &=&
(-1)^{\frac{1}{2}+\sigma_{\mbts{1}}+
\frac{1}{2}+\sigma_{\mbts{3}}}\,
{\cal F}^{-} (\bar{x}_1,x_2;\,\bar{x}_3,x_4)\,.
\label{dffchh}
\eea
The additional restoring amplitude ${\cal F}^{\,rest\,(c,c')}$
is introduced in Eq.~(\ref{dffcc}) for the purpose
of ``forced consistency'' and will be specified
in the following.

Making use of the definitions (\ref{inthfb}),
(\ref{projph})--(\ref{dffchh}) one can rewrite the
Eq.~(\ref{lrmeq}) in the channel form
\bea
\Lambda^{(c)} (x_1,x_2;\,\omega) &=&
\Lambda^{0\,(c)} (x_1,x_2;\,\omega) - \sum_{c'c''}
\int dx_3\,dx_4\,dx_5\,dx_6\,
A^{(c,c')} (x_1,x_2;\,x_3,x_4;\,\omega)
\nonumber\\
&\times& {\cal F}^{(c',c'')} (x_3,x_4;\,x_5,x_6)\,
\Lambda^{(c'')} (x_5,x_6;\,\omega)\,,
\label{lrmeqc}
\eea
where summation is performed over all the channels.

\section{EQUATION FOR PARTIAL COMPONENTS OF THE LRM\\
         IN CASE OF NEUTRAL EXCITATIONS \label{sect4}}

In the present paper we solve the LRM equation (\ref{lrmeqc})
for neutral excitations. To separate the angular dependence
in a spherically symmetric system we use decompositions
in terms of spherical tensor operators. In particular,
it is assumed that the operator of the renormalized external
field in the Eqs. (\ref{dflrm0}), (\ref{dfexfr}), (\ref{dfpol3})
has the form
\be
\tilde{V}^0 (x_1,x_2) = \tilde{V}^0_{JM} (x_1,x_2) =
\delta (\bfr_1 - \bfr_2)\,
\delta_{\tau_{\mbts{1}},\,\tau_{\mbts{2}}}\,
\sum_{LS} \tilde{V}^0_{JLS\tau_{\mbts{1}}} (r_1)\,
T_{JLSM} (\bfn_1)_{\sigma_{\mbts{1}}\sigma_{\mbts{2}}}\,,
\label{efpc}
\ee
where
\be
T_{JLSM} (\bfn)_{\sigma_{\mbts{1}}\sigma_{\mbts{2}}} =
\sum_{m\mu} (LmS\mu | JM)\, Y_{Lm} (\bfn)
(\sigma_{S\mu})_{\sigma_{\mbts{1}}\sigma_{\mbts{2}}}\,,
\label{dftjls1}
\ee
$\sigma_{S\mu}$ are the Pauli matrices in the tensor
representation:
\be
(\sigma_{S\mu})_{\sigma_{\mbts{1}}\sigma_{\mbts{2}}} =
\sqrt{2\,(2\,S+1)}\,(-1)^{\frac{1}{2}-\sigma_{\mbts{1}}}
\left(
\begin{array}{ccc}
\frac{1}{2} & \frac{1}{2} & S \\
\sigma_2    & -\sigma_1   & \mu \\
\end{array}
\right)\,.
\label{pmatr}
\ee
For electric multipole excitations we use the standard ansatz:
\be
\tilde{V}^0_{JLS\tau} (r) = \delta_{JL}\,\delta_{S\,0}\,r^L\,
e^{(L)}_{\tau}\,,
\label{efpce}
\ee
where $e^{(L)}_{\tau}$ is an effective charge
in the center-of-mass reference frame:
$\,e^{(L)}_n = Z\,(- A^{-1})^L$,
$e^{(L)}_p = (1 - A^{-1})^L + (Z - 1)\,(- A^{-1})^L$.

Let us denote in accordance with Eqs. (\ref{dflrm0}),
(\ref{dflrm0c}), (\ref{dfacph}), and (\ref{efpc}):
\be
\Lambda^{0\,(c)}_{JM} (x_1,x_2;\,\omega) =
-\int dx_3\,dx_4\,A^{(c,ph)} (x_1,x_2;\,x_3,x_4;\,\omega)\,
\tilde{V}^0_{JM} (x_4,x_3)\,.
\label{dflrm0jm}
\ee
Let $\Lambda^{(c)}_{JM}$ be the solution of the
Eq.~(\ref{lrmeqc}) with
$\Lambda^{0\,(c)} = \Lambda^{0\,(c)}_{JM}$. In this case
the partial components of the LRM $\Lambda^{(c)}_{JM}$
and of the related quantities are defined as
\bea
\Lambda^{(c)}_{JLS\tau} (r;\,\omega) &=&
\int d\bfn\,dx_1\,dx_2\,
T^{\,\dag}_{JLSM} (\bfn)_{\sigma_{\mbts{1}}\sigma_{\mbts{2}}}\,
\delta (x_1,x_2;\,\bfr,\tau)\,
\Lambda^{(c)}_{JM} (x_1,x_2;\,\omega)\,,
\label{dflrmj}\\
\Lambda^{0\,(c)}_{JLS\tau} (r;\,\omega) &=&
\int d\bfn\,dx_1\,dx_2\,
T^{\,\dag}_{JLSM} (\bfn)_{\sigma_{\mbts{1}}\sigma_{\mbts{2}}}\,
\delta (x_1,x_2;\,\bfr,\tau)\,
\Lambda^{0\,(c)}_{JM} (x_1,x_2;\,\omega)\,,
\label{dflrm0j}
\eea
\bea
A^{J\,(c,c')}_{LS\tau,L'S'\tau'} (r, r';\,\omega) &=&
\delta_{\tau,\tau'}\!\!
\int d\bfn\,d\bfn'\,dx_1\,dx_2\,dx_3\,dx_4\,
\delta (x_1,x_2;\,\bfr,\tau)\,\delta (x_3,x_4;\,\bfr',\tau')\,
\nonumber\\
&\times&
T^{\,\dag}_{JLSM} (\bfn)_{\sigma_{\mbts{1}}\sigma_{\mbts{2}}}\,
A^{(c,c')} (x_1,x_2;\,x_3,x_4;\,\omega)\,
T_{JL'S'M} (\bfn')_{\sigma_{\mbts{4}}\sigma_{\mbts{3}}}\,,
\label{ajls1}
\eea
where
\bea
T^{\,\dag}_{JLSM} (\bfn)_{\sigma_{\mbts{1}}\sigma_{\mbts{2}}}
&=& (-1)^{J+L+S+M}\,
T_{JLS-M} (\bfn)_{\sigma_{\mbts{1}}\sigma_{\mbts{2}}}\,,
\label{tjlsmc}\\
\delta (x_1,x_2;\,\bfr,\tau) &=&
\delta_{\tau_{\mbts{1}},\tau}\,\delta_{\tau_{\mbts{2}},\tau}\,
\delta (\bfr_1 - \bfr)\,\delta (\bfr_2 - \bfr)\,.
\label{dfdxrt}
\eea

Further, we assume that the effective interaction
is determined by the following decomposition:
\bea
{\cal F}^{(c,c')} (x_1,x_2;\,x_3,x_4) &=&
\delta (\bfr_1 - \bfr_2)\,\delta (\bfr_3 - \bfr_4)\,
\delta_{\tau_{\mbts{1}},\,\tau_{\mbts{2}}}\,
\delta_{\tau_{\mbts{3}},\,\tau_{\mbts{4}}}\,
\nonumber\\
&\times&
\sum_{LSL'S'JM}
T_{JLSM} (\bfn_1)_{\sigma_{\mbts{2}}\sigma_{\mbts{1}}}\,
T^{\,\dag}_{JL'S'M}
(\bfn_3)_{\sigma_{\mbts{3}}\sigma_{\mbts{4}}}
\nonumber\\
&\times&
{\cal F}^{J\,(c,c')}_{LS\tau_{\mbts{1}},\,L'S'\tau_{\mbts{3}}}
(r_1,r_3)\,,
\label{intdec}
\eea
where
\bea
{\cal F}^{J\,(c,c')}_{LS\tau,\,L'S'\tau'} (r,r') &=&
\delta_{c,c'}\,\delta_{LL'}\,\delta_{SS'}\,
\frac{\delta(r-r')}{rr'}\,
{\cal F}^{(c)}_{S,\,\tau \tau'} (r)
\nonumber\\
&+& \delta_{LL'}\,\delta_{L1}\,\delta_{J1}
\sum_{k=1}^2 \vk_{\,k}\,F^{(c)}_{S\tau,\,k} (r)\,
F^{(c')}_{S'\tau',\,k} (r')\,,
\label{intjls}\\
{\cal F}^{(ph)}_{S,\,\tau \tau'} &=&
C_0\,\bigr(\,\delta_{S0}\,\bigr[\,
f + (2\,\delta_{\tau,\tau'} - 1)\,f'\bigl] +
\delta_{S1}\,\bigr[\,
g + (2\,\delta_{\tau,\tau'} - 1)\,g'\bigl]\,\bigl)\,,
\label{intphst}\\
{\cal F}^{(pp)}_{S,\,\tau \tau'} &=&
{\cal F}^{(hh)}_{S,\,\tau \tau'} \,=\,
\delta_{S0}\,\delta_{\tau,\tau'}\,{\cal F}^{\,\xi}\,.
\label{intppst}
\eea

In Eq.~(\ref{intjls}) the first term
represents Landau-Migdal zero-range interaction of the TFFSPC.
In the standard parametrization,
the quantities $C_0$, $g$, and $g'$ in Eq.~(\ref{intphst})
are constants. The functions $f(r)$ and $f'(r)$ are
determined by the parameters
$f{\vphantom{f'}}_{ex}$, $f{\vphantom{f'}}_{in}$, $f'_{ex}$,
$f'_{in}$, and by the nuclear density in the ground state
$\rho_0(r)$ by means of the ansatz:
\be
f(r) = f_{ex} + (f_{in} - f_{ex})\,\rho_0(r)/\rho_0(0)\,,
\label{parfis}
\ee
and analogously for $f'(r)$.
For the interaction in the pp and hh channels we have:
\be
{\cal F}^{\,\xi} = C_0/\ln (c_p/\xi)\,,
\label{parfxi}
\ee
where $c_p$ is a constant, $\xi$ is an energy cutoff
parameter.

The second term in the Eq.~(\ref{intjls}) is a correction
corresponding to the amplitude ${\cal F}^{\,rest\,(c,c')}$
in the Eq.~(\ref{dffcc}).
The similar correction arises in the calculational
scheme with ``forced consistency'' which was developed in
Ref.~\cite{KLLT} to obtain the spurious isoscalar $1^-$
state at zero energy in the non-self-consistent approach
for the case when only ph channel is taken into account.
Here the straightforward generalization of this scheme
is presented in which the pp and hh channels
are also included. Making use of the same method as in
Ref.~\cite{KLLT} we obtain:
\be
F^{(c)}_{S\tau,\,1} (r) = \delta_{c,ph}\,\delta_{S0}\,
\frac{dU^{\tau}(r)}{dr}\,,\qquad
F^{(c)}_{S\tau,\,2} (r) = \sum_{\tau'}
{\cal F}^{(c)}_{S,\,\tau \tau'} (r)\,
Q^{(c)}_{S\tau'} (r)\,,
\label{frc1}
\ee
\be
\vk^{\,-1}_k = - \sum_{cS\tau} \int_0^{\infty}
dr\,r^2\,F^{(c)}_{S\tau,\,k} (r)\,Q^{(c)}_{S\tau} (r)\,,
\label{frc2}
\ee
where
\be
Q^{(c)}_{S\tau} (r) = \sum_{c'S'} \int_0^{\infty}
dr'\,r'^{\,2}\,
A^{1\,(c,c')}_{1S\tau,1S'\tau} (r, r';\,\omega_0)\,
F^{(c')}_{S'\tau,\,1} (r')\,,
\qquad \omega_0 \to 0\,.
\label{frc3}
\ee
In Eqs.~(\ref{frc1}), $U^{\tau}(r)$ is an auxiliary potential,
the well depth of which is chosen to satisfy the condition:
\be
\int_0^{\infty} dr\,r^3\,
\bigl[\,Z\,Q^{(ph)}_{0\,n} (r) - N\,Q^{(ph)}_{0\,p} (r)\,\bigr]
= 0\,,
\label{frc4}
\ee
which ensures the spurious state excitation probability
to be equal to zero.

Making use of the above definitions we can reduce the
Eq.~(\ref{lrmeqc}) in case of the spherically symmetric system
to the following equation for the partial components:
\bea
\Lambda^{(c)}_{JLS\tau} (r;\,\omega) &=&
\Lambda^{0\,(c)}_{JLS\tau} (r;\,\omega)
\nonumber\\
&-&
\sum_{L'S'\tau'c'} \sum_{L''S''\tau''c''}
\int_0^{\infty} dr'\,r'^{\,2}\,
\int_0^{\infty} dr''\,r''^{\,2}\,
A^{J\,(c,c')}_{LS\tau,L'S'\tau'} (r, r';\,\omega)
\nonumber\\
&\times&
{\cal F}^{J\,(c',c'')}_{L'S'\tau',\,L''S''\tau''} (r',r'')\,
\Lambda^{(c'')}_{JL''S''\tau''} (r'';\,\omega)\,,
\label{lrmeqpc}
\eea
where
\bea
\Lambda^{0\,(c)}_{JLS\tau} (r;\,\omega) &=& - \sum_{L'S'c'}
\int_0^{\infty} dr'\,r'^{\,2}\,
A^{J\,(c,c')}_{LS\tau,L'S'\tau} (r, r';\,\omega)\,
\tilde{V}^{0\,(c')}_{JL'S'\tau} (r')\,,
\label{dflrm0pc}\\
\tilde{V}^{0\,(c)}_{JLS\tau} (r) &=&
\delta^{\vphantom{0}}_{c,ph}
\tilde{V}^{0}_{JLS\tau} (r)\,.
\label{dfvzc}
\eea
In terms of the partial components
the Eq.~(\ref{dfpol3}) for the polarizability
takes the form:
\be
\Pi_J(\omega) = (2J+1) \sum_{LS\tau c} \int_0^{\infty}
dr\,r^2\,\tilde{V}^{0\,(c)}_{JLS\tau} (r)\,
\Lambda^{(c)}_{JLS\tau} (r;\,\omega)\,.
\label{dfpol4}
\ee

It is important to note that the amplitudes
$u^{\vphantom{*}}_{\lambda}$ and $v^{\vphantom{*}}_{\lambda}$,
which define the single-quasiparticle basis functions
according to Eq.~(\ref{uvpsisa}), have to be determined
from the solution of the gap equation with the same interaction
${\cal F}^{\,\xi}$ which enters the Eq.~(\ref{lrmeqpc})
for the LRM through the Eqs. (\ref{intjls}) and (\ref{intppst}).
Namely, we have:
\be
u^{\vphantom{*}}_{\lambda} = \sqrt{\frac{1}{2}\,\left( 1+
\frac{\ve^{\vphantom{2}}_{\lambda} -
\mu_{\tau_{\mbts{$\lambda$}}}}
{E_{\lambda}} \right)}
\,,\qquad
v^{\vphantom{*}}_{\lambda} = \mbox{sgn}(\Delta_{\lambda})\,
\sqrt{\frac{1}{2}\,\left( 1-
\frac{\ve^{\vphantom{2}}_{\lambda} -
\mu_{\tau_{\mbts{$\lambda$}}}}
{E_{\lambda}} \right)}\,,
\label{uvfor}
\ee
where $\ve^{\vphantom{2}}_{\lambda}$ is the eigenvalue of the
single-particle Hamiltonian corresponding to the eigenfunction
$\vphi^{\vphantom{*}}_{\lambda}(x)$, $\;\mu_{\tau}$
is the chemical potential,
\be
E_{\lambda} = \sqrt{ (
\ve^{\vphantom{2}}_{\lambda} -
\mu_{\tau_{\mbts{$\lambda$}}} )^2
+ \Delta_{\lambda}^2}\,.
\label{dfqpen}
\ee
The values of the energy gap $\Delta_{\lambda}$ are determined
within the BCS approximation from the equation:
\be
\Delta_{\lambda} = - \sum_{(\lambda')}
\,\frac{2j_{\lambda'}+1}{4\pi}\,
{\cal F}^{\,\xi}_{(\lambda \lambda')}\,
\frac{\Delta_{\lambda'}}{2E_{\lambda'}}\,,
\label{gapeq}
\ee
where
\be
{\cal F}^{\,\xi}_{(\lambda \lambda')} =
\delta_{\tau_{\vphantom{\lambda'}\lambda}\,,\,
\tau_{\vphantom{\lambda'}\lambda'}}\,
\int_0^{\infty} dr\,r^2\,
R^{\,2}_{(\lambda)} (r)\,R^{\,2}_{(\lambda')} (r)\,
{\cal F}^{\,\xi}(r)\,.
\label{fxill}
\ee

\section{CORRELATED PROPAGATOR\\
         IN THE COORDINATE REPRESENTATION \label{cpcr}}

Let us obtain an explicit formula for the partial components of
the correlated propagator in terms of the reduced matrix elements.
First of all, notice that the following relations are fulfilled
for the matrix elements $A_{12,34}(\omega)$ entering
Eq.~(\ref{propcr}) in case of the spherically symmetric system:
\bea
A_{12,34}(\omega) &=& \sum_{JM} (2J+1)\,
A^J_{[12,34]}(\omega)
\nonumber\\
&\times&
(-1)^{j_{\mbts{2}}-m_{\mbts{2}}}\,
\left(
\begin{array}{ccc}
j_1 & j_2 & J \\ m_1 & -m_2 & M \\
\end{array}
\right)\,
(-1)^{j_{\mbts{4}}-m_{\mbts{4}}}\,
\left(
\begin{array}{ccc}
j_3 & j_4 & J \\ m_3 & -m_4 & M \\
\end{array}
\right)\,,
\label{arme1}\\
A^J_{[12,34]}(\omega) &=& \sum_{m_1 m_2 m_3 m_4} (2J+1)\,
A_{12,34}(\omega)\,
\nonumber\\
&\times&
(-1)^{j_{\mbts{2}}-m_{\mbts{2}}}\,
\left(
\begin{array}{ccc}
j_1 & j_2 & J \\ m_1 & -m_2 & M \\
\end{array}
\right)\,
(-1)^{j_{\mbts{4}}-m_{\mbts{4}}}\,
\left(
\begin{array}{ccc}
j_3 & j_4 & J \\ m_3 & -m_4 & M \\
\end{array}
\right)\,,
\label{arme2}
\eea
where $1 = \{[1], m_1 \}$, $[1] = \{(1), \eta_1 \}$,
$(1) = (\lambda_1) = \{\tau_1, n_1, l_1, j_1 \}$.
Notice that with these abbreviated notations we have:
$u^{\vphantom{(2)}}_{(1)}
=u^{\vphantom{(2)}}_{\lambda_{\mbts{1}}}$,
$\;v^{\vphantom{(2)}}_{(1)}
=v^{\vphantom{(2)}}_{\lambda_{\mbts{1}}}$,
$\;\ve^{\vphantom{(2)}}_{(1)}
=\ve^{\vphantom{(2)}}_{\lambda_{\mbts{1}}}$,
$\;E^{\vphantom{(2)}}_{(1)}
=E^{\vphantom{(2)}}_{\lambda_{\mbts{1}}}$.
We will use the antisymmetric form of equations for
the correlated propagator (see Ref.~\cite{T05} for details)
which allows to reduce the dimensions of matrices entering
these equations. In this case the following relations
are fulfilled:
\bea
A^J_{[12,34]}(\omega)
&=& - \eta_{\mbss{1}}\,\eta_{\mbss{2}}\,
(-1)^{J+l_{\mbts{1}}-l_{\mbts{2}}+j_{\mbts{1}}-j_{\mbts{2}}}\,
A^J_{[\bar{2}\bar{1},34]}(\omega)
\nonumber\\
&=& - \eta_{\mbss{3}}\,\eta_{\mbss{4}}\,
(-1)^{J+l_{\mbts{3}}-l_{\mbts{4}}+j_{\mbts{3}}-j_{\mbts{4}}}\,
A^J_{[12,\bar{4}\bar{3}]}(\omega)\,,
\label{asprred}
\eea
where $[\bar{1}] = \{(1), -\eta_1 \}$.

Making use of Eqs. (\ref{uvpsisa}), (\ref{relphi}),
(\ref{dfmo}), and (\ref{dfpsit}) it is easy to show
that the functions $\tilde{\psi}_1(y)$ entering
Eq.~(\ref{propcr}) obey the equalities:
\be
\tilde{\psi}^{\vphantom{(2)}}_1(x;+) =
w^{\vphantom{(2)}}_1\,
\vphi^{\vphantom{(2)}}_{\lambda_{\mbts{1}}}(x)\,,
\qquad
(-1)^{\frac{1}{2}+\sigma}\,
\tilde{\psi}^{\vphantom{(2)}}_1(\bar{x};-) =
\eta^{\vphantom{(2)}}_1\,w^{\vphantom{(2)}}_{\bar{1}}\,
\vphi^{\vphantom{(2)}}_{\lambda_{\mbts{1}}}(x)\,,
\label{relpsi}
\ee
where
\be
w^{\vphantom{(2)}}_1 = w^{\vphantom{(2)}}_{[1\,]} =
\delta^{\vphantom{(2)}}_{\eta_{\mbts{1}},+1}\,
u^{\vphantom{(2)}}_{(1)} +
\delta^{\vphantom{(2)}}_{\eta_{\mbts{1}},-1}\,
v^{\vphantom{(2)}}_{(1)}\,.
\label{dfw1}
\ee
Let us define the reduced matrix elements of the operator
$T_{JLSM}$ by the relation:
\be
\langle \,j_1 l_1 m_1\,|\,T_{JLSM}\,|\,j_2 l_2 m_2\,\rangle =
(-1)^{j_{\mbts{1}}-m_{\mbts{1}}}\,
\left(
\begin{array}{ccc}
j_1 & j_2 & J \\ m_1 & -m_2 & -M \\
\end{array}
\right)\,
\langle \,j_1 l_1\,||\,T_{JLS}\,||\,j_2 l_2\, \rangle \,.
\label{trme}
\ee
In the explicit form we have:
\bea
\langle \,j_1 l_1\,||\,T_{JLS}\,||\,j_2 l_2\, \rangle &=&
\frac{1}{2}\,
\bigl[ 1+(-1)^{L+l_{\mbts{1}}+l_{\mbts{2}}} \bigr]\,
(-1)^{S+j_{\mbts{2}}-\frac{1}{2}}
\nonumber\\
&\times&
\sqrt{\frac{(2J+1)\,(2L+1)\,(2j_1+1)\,(2j_2+1)}{4\pi}}\,
\left(
\begin{array}{ccc}
j_1 & j_2 & J \\ \frac{1}{2} & -\frac{1}{2} & 0 \\
\end{array}
\right)
\nonumber\\
&\times&
\Biggl\{
\left(
\begin{array}{ccc}
J & L & S \\ 0 & 0 & 0 \\
\end{array}
\right)
+ \sqrt{\frac{S\,(S+1)}{J\,(J+1)}}\,
\left(
\begin{array}{ccc}
J & L & S \\ 1 & 0 & -1 \\
\end{array}
\right)
\nonumber\\
&\times&
\bigl[\,
(l_1 - j_1)\,(2j_1+1) +
(-1)^{J+L+S}\,(l_2 - j_2)\,(2j_2+1)
\,\bigr]\,
\Biggr\}\,.
\label{extrme}
\eea

Making use of Eqs.
(\ref{dfphil}), (\ref{dfsaf}), (\ref{propcr}),
(\ref{projph})--(\ref{projhh}),
(\ref{dfacph})--(\ref{dfachh}),
(\ref{arme1}), (\ref{asprred}), (\ref{relpsi}), (\ref{trme}),
we obtain from the Eq.~(\ref{ajls1}) the following ansatz:
\bea
A^{J\,(c,c')}_{LS\tau,L'S'\tau'} (r, r';\,\omega) &=&
\bigl(\,1 + \delta^{\vphantom{(+)}}_{c'\!,\,ph}\,\bigr)\,
\delta_{\tau,\tau'} \sum_{[1234]}
\delta_{\tau_{\mbts{1}},\,\tau}\,
\delta_{\tau_{\mbts{2}},\,\tau}\,
\delta_{\tau_{\mbts{3}},\,\tau'}\,
\delta_{\tau_{\mbts{4}},\,\tau'}\,
\theta^{\vphantom{(+)}}_{(21)}\,
\theta^{\vphantom{(+)}}_{(43)}
\nonumber\\
&&\times\,
A^{J\,(c,c')\,LS,L'S'}_{[12,34]} (r, r';\,\omega)\,,
\label{ajls2}
\eea
where
\bea
A^{J\,(c,c')\,LS,L'S'}_{[12,34]} (r, r';\,\omega) &=&
R^{\vphantom{(J)}}_{(1)}(r)\,R^{\vphantom{(J)}}_{(2)}(r)\,
R^{\vphantom{(J)}}_{(3)}(r')\,R^{\vphantom{(J)}}_{(4)}(r')\,
T^{(J)\,LS,L'S'}_{(12,34)}
\nonumber\\
&&\times
\bigl(\alpha^c_{\vphantom{\bar{1}}[12]} -
\eta_{\mbss{1}}\eta_{\mbss{2}}\,
(-1)^{S}\alpha^c_{[\bar{2}\bar{1}]}\bigr)
\,
A^J_{[12,34]}(\omega)\,
\nonumber\\
&&\times
\bigl(\alpha^{c'}_{\vphantom{\bar{1}}[34]} -
\eta_{\mbss{3}}\eta_{\mbss{4}}\,
(-1)^{S'}\alpha^{c'}_{[\bar{4}\bar{3}]}\bigr)\,,
\label{ajls3}
\eea
\be
T^{(J)\,LS,L'S'}_{(12,34)} = \frac{1}{2J+1}\,
\langle \,j_2 l_2\,||\,T_{JLS}\,||\,j_1 l_1\, \rangle \,
\langle \,j_4 l_4\,||\,T_{JL'S'}\,||\,j_3 l_3\, \rangle \,,
\label{dftjls2}
\ee
\be
\alpha^{ph}_{[12]} = w_{\mbss{1}}\,w_{\mbss{2}}\,,\qquad
\alpha^{pp}_{[12]} =
\eta_{\mbss{2}}\,w_{\mbss{1}}\,w^{\vphantom{h_p}}_{\bar{2}}\,,
\qquad
\alpha^{hh}_{[12]} =
\eta_{\mbss{1}}\,w^{\vphantom{h_p}}_{\bar{1}}\,w_{\mbss{2}}\,.
\label{dfalc}
\ee

Theoretically, summation in the Eq.~(\ref{ajls2}) is supposed
to be over complete ordered set of the states
forming the doubled configuration space.
This summation is facilitated due to the symmetry
defined by the Eq.~(\ref{asprred}).
The order-bounding factors are defined as follows:
$\,\theta^{\vphantom{(+)}}_{(21)} = 1\,$
if the ordinal number of the state $(1)$ is lesser than
the number of $(2)$ $\,[(1) < (2)]$,
$\,\theta^{\vphantom{(+)}}_{(21)} = \frac{1}{2}\,$ if $\,(1) = (2)$,
$\,\theta^{\vphantom{(+)}}_{(21)} = 0\,$ if $\,(1) > (2)$.

However, in our calculational scheme, summation in the
Eq.~(\ref{ajls2}) is restricted by the discrete and quasidiscrete
states entering a valence zone near the Fermi level.
It is supposed that the remaining part of the sum can be
approximated fairly well by the RPA-like propagator
$A^{\,cont}$, which contains transitions from the
quasiparticle levels to the continuum.
Thus, we use the scheme which is analogous to the
ones described in Refs.~\cite{KL04,KLLT}. According to this
scheme the total correlated propagator is represented
as a sum of two terms:
\be
A^{J\,(c,c')}_{LS\tau,L'S'\tau'} (r, r';\,\omega) =
A^{J\,(c,c')\,cont}_{LS\tau,L'S'\tau'} (r, r';\,\omega) +
A^{J\,(c,c')\,disc}_{LS\tau,L'S'\tau'} (r, r';\,\omega)\,,
\label{ajls4}
\ee
where
\bea
&&A^{J\,(c,c')\,cont}_{LS\tau,L'S'\tau'} (r, r';\,\omega) =
- \,\delta_{c,c'}\,\delta_{c,ph}\,\delta_{\tau,\tau'}
\bigg\{ \sum_{j_1 l_1}\,\sum_{(2)}^{disc}
\delta_{\tau_{\mbts{2}},\,\tau}\,v^2_{(2)}\,
R^{\vphantom{(2)}}_{(2)}(r)\,R^{\vphantom{(2)}}_{(2)}(r')
\nonumber\\
&&\times\,
T^{(J)\,LS,L'S'}_{(12,12)}
\bigl[\,
\tilde{G}^{\,nor}_{j_{\mbss{1}} l_{\mbss{1}} \tau}
(r,r';\,\mu_{\tau} - E_{(2)} + \omega) + (-1)^{S+S'}
\tilde{G}^{\,nor}_{j_{\mbss{1}} l_{\mbss{1}} \tau}
(r,r';\,\mu_{\tau} - E_{(2)} - \omega)\, \bigr]
\nonumber\\
&&- \sum_{(12)}^{disc}
\delta_{\tau_{\mbts{1}},\,\tau}\,
\delta_{\tau_{\mbts{2}},\,\tau}\,
\theta^{\vphantom{(+)}}_{(21)}\,
R^{\vphantom{(2)}}_{(1)}(r)\,R^{\vphantom{(2)}}_{(2)}(r)\,
R^{\vphantom{(2)}}_{(1)}(r')\,R^{\vphantom{(2)}}_{(2)}(r')
\nonumber\\
&&\times\,
T^{(J)\,LS,L'S'}_{(12,12)}
\bigg[\,
\frac{v^2_{(2)}}{\omega + \mu_{\tau} -
E^{\vphantom{(2)}}_{(2)} - \ve^{\vphantom{(2)}}_{(1)}} -
\frac{v^2_{(1)}}{\omega - \mu_{\tau} +
E^{\vphantom{(2)}}_{(1)} + \ve^{\vphantom{(2)}}_{(2)}}
\nonumber\\
&&+\;\,\,(-1)^{S+S'}
\bigg(\,
\frac{v^2_{(1)}}{\omega + \mu_{\tau} -
E^{\vphantom{(2)}}_{(1)} - \ve^{\vphantom{(2)}}_{(2)}} -
\frac{v^2_{(2)}}{\omega - \mu_{\tau} +
E^{\vphantom{(2)}}_{(2)} + \ve^{\vphantom{(2)}}_{(1)}}\,
\bigg) \bigg] \bigg\}\,,
\label{acont}
\eea
\bea
&&A^{J\,(c,c')\,disc}_{LS\tau,L'S'\tau'} (r, r';\,\omega) =
\delta_{c,c'}\,\delta_{c,ph}\,\delta_{\tau,\tau'}
\sum_{(12)}^{disc}
\delta_{\tau_{\mbts{1}},\,\tau}\,
\delta_{\tau_{\mbts{2}},\,\tau}\,
\theta^{\vphantom{(+)}\,b}_{(1)}\,
[1 - \theta^{\vphantom{(+)}\,b}_{(2)}]\;T^{(J)\,LS,L'S'}_{(12,12)}
\nonumber\\
&&\times\;
R^{\vphantom{(2)}}_{(1)}(r)\,R^{\vphantom{(2)}}_{(2)}(r)\,
R^{\vphantom{(2)}}_{(1)}(r')\,R^{\vphantom{(2)}}_{(2)}(r')\,
\big[\,v^2_{(1)}\,(1 - v^2_{(2)}) +
(-1)^{S+S'} v^2_{(2)}\,(1 - v^2_{(1)})\,\big]
\vphantom{\frac{1}
{\omega + E^{\vphantom{(2)}}_{(1)} + E^{\vphantom{(2)}}_{(2)}}}
\nonumber\\
&&\times\,
\bigg[\,
\frac{1}
{\omega + E^{\vphantom{(2)}}_{(1)} + E^{\vphantom{(2)}}_{(2)}}
- \frac{(-1)^{S+S'}}
{\omega - E^{\vphantom{(2)}}_{(1)} - E^{\vphantom{(2)}}_{(2)}}
\bigg]
\nonumber\\
&&\mathop{+}\,
\bigl(\,1 + \delta^{\vphantom{(+)}}_{c'\!,\,ph}\,\bigr)\,
\delta_{\tau,\tau'} \sum_{[1234]}^{wind}
\delta_{\tau_{\mbts{1}},\,\tau}\,
\delta_{\tau_{\mbts{2}},\,\tau}\,
\delta_{\tau_{\mbts{3}},\,\tau'}\,
\delta_{\tau_{\mbts{4}},\,\tau'}\,
\theta^{\vphantom{(+)}}_{(21)}\,
\theta^{\vphantom{(+)}}_{(43)}\,
A^{J\,(c,c')\,LS,L'S'}_{[12,34]} (r, r';\,\omega)\,.
\label{adisc}
\eea
In Eq.~(\ref{acont}), $\tilde{G}^{\,nor}_{jl\tau} (r,r';\,\ve)$
is the partial component of the normal single-particle GF
without pairing. It is calculated via the regular and irregular
solutions of the Schr\"odinger equation that allows to take
into account continuum completely on the RPA level
(see, e.g., Ref.~\cite{SB}).
The relation between the Eqs. (\ref{ajls4})--(\ref{adisc})
and the initial Eqs. (\ref{ajls2}), (\ref{ajls3})
is determined by the formal decomposition:
\be
\tilde{G}^{\,nor}_{jl\tau} (r,r';\,\ve) = \sum_{(1)}
\delta_{j_{\mbts{1}},\,j}\,
\delta_{l_{\mbts{1}},\,l}\,
\delta_{\tau_{\mbts{1}},\,\tau}\,
\frac{R^{\vphantom{(2)}}_{(1)}(r)\,
R^{\vphantom{(2)}}_{(1)}(r')}
{\ve - \ve^{\vphantom{(2)}}_{(1)}}\,.
\label{gjlex}
\ee
In Eq.~(\ref{adisc}), $\,\theta^{\vphantom{(+)}\,b}_{(1)}$ are
the bounding factors defined as:
$\,\theta^{\vphantom{(+)}\,b}_{(1)} = 1\,$ if
$\,(1) < (\lambda_b)$,
$\,\theta^{\vphantom{(+)}\,b}_{(1)} = 0\,$ if
$\,(1) \geqslant (\lambda_b)$,
where $\lambda_b$ is the bottom level of the valence zone
which coincides in our calculations with ``pairing window'',
i.e. with zone where the energy gap is not equal to zero
(see Sec.~\ref{results} for details).
The superscript ``$disc$''
in the $\sum$ symbols means summation over
all discrete and quasidiscrete states of the single-particle
spectrum. The superscript ``$wind$'' means summation over
the discrete and quasidiscrete states only inside
the ``pairing window'', i.e. the same summation as in
the Eq.~(\ref{gapeq}).

\section{THE AMPLITUDE OF THE QUASIPARTICLE-PHONON\\
         INTERACTION AND ITS MATRIX ELEMENTS \label{qpi}}

Matrix elements of the amplitude of quasiparticle-phonon
interaction $g^m_{12}$ belong to the basic quantities entering
formulas for the correlated propagator of the QTBA
(see Ref.~\cite{T05} and Sec.~\ref{proprme} of the present paper).
In accordance with the general definition of this amplitude
in the microscopic approach we have:
\be
g^m (y_1,y_2) = \int dy_3\,dy_4\,
{\cal F} (y_1,y_2;\,y_3,y_4)\,
\rho^{m0} (y_3,y_4)\,,
\label{dfgcr}
\ee
where the interaction amplitude ${\cal F}$
is defined by the Eq.~(\ref{inthfb}), $\rho^{m0}$ is
the transition amplitude:
\be
\rho^{m0} (y_1,y_2) =
\langle \,m\,|\,b^{\dag} (y_1)\, b (y_2)\,|\, 0\,\rangle\,.
\label{dftdc}
\ee
The operators $b(y)=b(x,\chi)$ are defined by the relations:
\be
b(x,+) = a(x)\,,\qquad b(x,-) = a^{\dag}(x)\,,
\label{dfb}
\ee
where $a^{\dag}(x)$ and $a(x)$ are creation and annihilation
operators of particles.

As a first step we consider the coupling of the
quasiparticles with core excitations only in the ph channel.
So, substituting Eq.~(\ref{inthfb}) into Eq.~(\ref{dfgcr}),
we keep only the contributions of the amplitude
${\cal F}^{+}$. Taking into account the symmetry:
$\rho^{m0} (y_1,y_2) = -\rho^{m0} (\bar{y}_2,\bar{y}_1)$,
we obtain from Eqs. (\ref{inthfb}) and (\ref{dfgcr}):
\be
g^m (y_1,y_2) =
\delta_{\chi_{\mbts{1}},\,\chi_{\mbts{2}}}\,
\bigl[\,
\delta_{\chi_{\mbts{1}},\,+1}\,g^m (x_1,x_2) -
\delta_{\chi_{\mbts{1}},\,-1}\,g^m (x_2,x_1)\,\bigr]\,,
\label{gcrph}
\ee
where
\be
g^m (x_1,x_2) = \int dx_3\,dx_4\,
{\cal F}^{+} (x_1,x_2;\,x_3,x_4)\,
\rho^{m0} (x_3+,x_4+)\,.
\label{dfgcrx}
\ee

In what follows we assume that the function $g^m (x_1,x_2)$
has the form:
\be
g^m (x_1,x_2) = g^{qM} (x_1,x_2) =
\delta (\bfr_1 - \bfr_2)\,
\delta_{\tau_{\mbts{1}},\,\tau_{\mbts{2}}}\,\sum_{LS}
g^{qLS}_{\tau_{\mbts{1}}} (r_1)\,
T_{J_qLSM} (\bfn_1)_{\sigma_{\mbts{2}}\sigma_{\mbts{1}}}\,,
\label{gcrxd}
\ee
where we separate the projection of the total angular momentum
$M$ from the remaining phonon quantum numbers,
which are denoted by the index $q$ (including the total
angular momentum $J_q$). Notice that
Eq.~(\ref{gcrxd}) is in accordance with Eq.~(\ref{dfgcrx})
and with the decomposition of the effective interaction
within the TFFSPC (\ref{intdec}).

Let us denote
\bea
g^{qM}_{12} &=& \int dy_1\,dy_2\,
\tilde{\psi}^{\vphantom{*}}_1(y_1)\,
\tilde{\psi}^*_2(y_2)\,g^{qM} (y_1,y_2)\,,
\label{gmat1}\\
g^{qM}_{\lambda_{\mbts{1}} \lambda_{\mbts{2}}} &=&
\int dx_1\,dx_2\,
\vphi^{\vphantom{*}}_{\lambda_{\mbts{1}}} (x_1)\,
\vphi^*_{\lambda_{\mbts{2}}} (x_2)\,g^{qM} (x_1,x_2)\,.
\label{gmat2}
\eea
From Eqs. (\ref{dfphil}), (\ref{trme}), (\ref{gcrxd}),
and (\ref{gmat2}) it follows that
\be
g^{qM}_{\lambda_{\mbts{1}} \lambda_{\mbts{2}}} =
(-1)^{j_{\mbts{2}}-m_{\mbts{2}}}\,
\left(
\begin{array}{ccc}
j_2 & j_1 & J_q \\ m_2 & -m_1 & -M \\
\end{array}
\right)\,
\sum_S\,g^{qS}_{(12)}\,,
\label{gmat3}
\ee
where
\bea
g^{qS}_{(12)} &=& \sum_L\,g^{qLS}_{(12)}\,
\langle \,j_2 l_2\,||\,T_{J_qLS}\,||\,j_1 l_1\, \rangle \,,
\label{gmat4}\\
g^{qLS}_{(12)} &=&
\delta_{\tau_{\mbts{1}},\,\tau_{\mbts{2}}}
\int_0^{\infty} dr\,r^2\,R^{\vphantom{(2)}}_{(1)}(r)\,
R^{\vphantom{(2)}}_{(2)}(r)\,
g^{qLS}_{\tau_{\mbts{1}}}(r)\,.
\label{gmat5}
\eea
Making use of Eqs. (\ref{relpsi}), (\ref{gcrph}),
(\ref{gmat1})--(\ref{gmat3}), and the symmetry
of the operator $T_{JLSM}$ we obtain the following two main
formulas for the matrix elements of the amplitude
of quasiparticle-phonon interaction:
\be
g^{qM}_{12} =
(-1)^{j_{\mbts{2}}-m_{\mbts{2}}}\,
\left(
\begin{array}{ccc}
j_2 & j_1 & J_q \\ m_2 & -m_1 & -M \\
\end{array}
\right)\,
g^{q}_{[12]}\,,
\label{gmat6}
\ee
where
\be
g^{q}_{[12]} = \sum_S\,
\bigl(\,w^{\vphantom{(2)}}_1\,w^{\vphantom{(2)}}_2
- \eta^{\vphantom{(2)}}_1\,\eta^{\vphantom{(2)}}_2\,(-1)^S\,
w^{\vphantom{*}}_{\bar{1}}\,w^{\vphantom{*}}_{\bar{2}}\,
\bigl)\,g^{qS}_{(12)}\,.
\label{gmat7}
\ee

The Eq.~(\ref{gmat7}) can be easily generalized to take into
account the coupling of the quasiparticles with core excitations
in the pp and the hh channels. Proceeding in the same way,
we obtain:
\be
g^{q}_{[12]} = \sum_{LSc}\,
\bigl(\,\alpha^c_{\vphantom{\bar{1}}[12]} -
\delta^{\vphantom{(+)}}_{c,\,ph}\,
\eta_{\mbss{1}}\eta_{\mbss{2}}\,
(-1)^{S}\alpha^c_{[\bar{2}\bar{1}]}\,\bigr)\,
\langle \,j_2 l_2\,||\,T_{J_qLS}\,||\,j_1 l_1\, \rangle \,
g^{qLS(c)}_{(12)}\,,
\label{gmat7c}
\ee
where the amplitudes $\alpha^c_{\vphantom{\bar{1}}[12]}$
are defined by the Eqs.~(\ref{dfalc}),
\be
g^{qLS(c)}_{(12)} =
\delta_{\tau_{\mbts{1}},\,\tau_{\mbts{2}}}
\int_0^{\infty} dr\,r^2\,R^{\vphantom{(2)}}_{(1)}(r)\,
R^{\vphantom{(2)}}_{(2)}(r)\,
g^{qLS(c)}_{\tau_{\mbts{1}}}(r)\,.
\label{gmat5c}
\ee
Notice that in the Eqs. (\ref{gcrxd}) and (\ref{gmat5})
we have: $g^{qLS}_{\tau}(r) = g^{qLS(ph)}_{\tau}(r)$.
To determine the function $g^{qLS(c)}_{\tau}(r)$ entering
Eq.~(\ref{gmat5c}) the following method is possible.
In terms of the partial components the Eq.~(\ref{dfgcr}) reads:
\be
g^{qLS(c)}_{\tau}(r) = \sum_{\tau'}
{\cal F}^{(c)}_{S,\,\tau \tau'} (r)\,
\rho^{qLS(c)}_{\tau'}(r)\,,
\label{gfrho}
\ee
where functions ${\cal F}^{(c)}_{S,\,\tau \tau'} (r)$
are determined by the Eqs. (\ref{intphst}) and (\ref{intppst}),
$\rho^{qLS(c)}_{\tau}(r)$ is the partial component of
the transition amplitude, and the contribution of the
restoring amplitude in Eq.~(\ref{intjls}) is omitted.
From the spectral decomposition of
the partial component of the LRM one can find that up to
irrelevant constant phase factor the following equality
is fulfilled:
\be
\rho^{qLS(c)}_{\tau}(r) =
\sqrt{\frac{2J_q+1}{B(q)\!\uparrow}}\,
\int\limits_{C_q}
\frac{d\omega}{2 \pi i}\,
\Lambda^{(c)}_{J_{\mbts{$q$}}LS\tau} (r;\,\omega)\,,
\label{tdpc}
\ee
where $\Lambda^{(c)}_{JLS\tau}$ is a solution of
Eq.~(\ref{lrmeqpc}), $\,B(q)\!\uparrow \,= B(\,\mbox{g.s.} \to q)$
is reduced excitation probability which is determined by the formula
\be
B(q)\!\uparrow \,=
\int\limits_{C_q}
\frac{d\omega}{2 \pi i}\,\Pi_{J_{\mbts{$q$}}} (\omega)\,.
\label{dfbqj}
\ee
The polarizability $\Pi_J (\omega)$ in Eq.~(\ref{dfbqj})
is determined by Eq.~(\ref{dfpol4}).
In the Eqs. (\ref{tdpc}) and (\ref{dfbqj})
the integration is performed along the contour $C_q$
in the complex energy plane which encloses the poles
of the functions
$\Lambda^{(c)}_{J_{\mbts{$q$}}LS\tau} (r;\,\omega)$
and $\Pi_{J_{\mbts{$q$}}} (\omega)$ corresponding to the
phonon energy $\omega_q$.
This method automatically ensures correct normalization of
the functions $\rho^{qLS(c)}_{\tau}(r)$ and
$g^{qLS(c)}_{\tau}(r)$. However, to determine these functions
in case of the excitations in
the pp and the hh channels one has to solve
Eq.~(\ref{lrmeqpc}) for the different external fields
$\tilde{V}^{0\,(c)}_{JLS\tau} (r)$ which take non-zero values
for $c = pp\,$ and $c = hh\,$ in contrast to the definition
(\ref{dfvzc}).

Notice that in order to calculate the function
$g^{qLS(c)}_{\tau}(r)$ by means of the formulas
(\ref{gfrho})--(\ref{dfbqj}) it is not necessary to know
the exact value of the phonon energy. It is sufficient
that the pole $\omega=\omega_q$ (and only this pole)
would be inside the contour $C_q$. The exact value
of $\omega_q$ can be determined by the same method
with making use of the formula
\be
\omega_q =
\frac{1}{B(q)\!\uparrow}
\int\limits_{C_q}
\frac{d\omega}{2 \pi i}\,\omega\,\Pi_{J_{\mbts{$q$}}} (\omega)\,.
\label{dfeqj}
\ee

Another method to determine matrix elements
$g^{q}_{[12]}$ consists in solving QRPA equation for
the transition amplitudes making use of the basis restricted
by discrete and quasidiscrete single-particle states
entering a valence zone near the Fermi level.
In terms of the reduced matrix elements this equation reads:
\be
\bigl(\,\omega_q - \eta\,
\bigl[\,E_{(1)} + E_{(2)}\bigr] \bigr)
\rho^{q}_{(12)\eta} =
\sum_{\eta'} \sum^{wind}_{(34)}\eta\,
{\cal F}^{J_q}_{(12)\eta,\,(34)\eta'}\,
\theta^{\vphantom{(+)}}_{(43)}\,
\rho^{q}_{(34)\eta'}\,,
\label{dbas1}
\ee
where
\bea
{\cal F}^{J}_{(12)\eta,\,(34)\eta'} &=&
\sum_{\eta_{\mbts{1}}\eta_{\mbts{2}}\eta_{\mbts{3}}\eta_{\mbts{4}}}
\delta^{\vphantom{(+)}}_{\eta_{\mbts{1}},\,\eta\vphantom{\eta'}}\,
\delta^{\vphantom{(+)}}_{\eta_{\mbts{2}},-\eta\vphantom{\eta'}}\,
\delta^{\vphantom{(+)}}_{\eta_{\mbts{3}},\,\eta'}\,
\delta^{\vphantom{(+)}}_{\eta_{\mbts{4}},-\eta'}\,
{\cal F}^{J}_{[12,34]}\,,
\label{dbas2}\\
{\cal F}^{J}_{[12,34]} &=& \frac{1}{2} \sum_{LSc}
\,\bigl(\alpha^c_{\vphantom{\bar{1}}[12]} -
\eta_{\mbss{1}}\eta_{\mbss{2}}\,
(-1)^{S}\alpha^c_{[\bar{2}\bar{1}]}\bigr)
\bigl(\alpha^c_{\vphantom{\bar{1}}[34]} -
\eta_{\mbss{3}}\eta_{\mbss{4}}\,
(-1)^{S}\alpha^c_{[\bar{4}\bar{3}]}\bigr)
\nonumber\\
&&\times
\bigl(\,1 + \delta^{\vphantom{(+)}}_{c\!,\,ph}\,\bigr)\,
T^{(J)\,LS,LS}_{(12,34)}\,
{\cal F}^{(c)S}_{(12,34)}\,,
\vphantom{\frac{A}{B_C}}
\label{dbas3}\\
{\cal F}^{(c)S}_{(12,34)} &=&
\delta_{\tau_{\mbts{1}},\,\tau_{\mbts{2}}}\,
\delta_{\tau_{\mbts{3}},\,\tau_{\mbts{4}}}
\int_0^{\infty}dr\,r^2\,
R^{\vphantom{(J)}}_{(1)}(r)\,R^{\vphantom{(J)}}_{(2)}(r)\,
R^{\vphantom{(J)}}_{(3)}(r)\,R^{\vphantom{(J)}}_{(4)}(r)\,
{\cal F}^{(c)}_{S,\,\tau_{\mbts{1}} \tau_{\mbts{3}}} (r)\,.
\label{dbas4}
\eea
The quantities entering these formulas are defined by the
Eqs. (\ref{intphst}), (\ref{intppst}),
(\ref{dftjls2}), (\ref{dfalc}).
Notice that QRPA equation (\ref{dbas1}) includes contributions
of all the channels: ph, pp, and hh.

Normalization condition for the transition amplitudes
$\rho^{q}_{(12)\eta}$ has the form
\be
\sum_{\eta} \sum^{wind}_{(12)}
\theta^{\vphantom{(+)}}_{(21)}\,\eta\,
\bigl| \rho^{q}_{(12)\eta} \bigr|^2
= 2J_q + 1\,.
\label{dbas5}
\ee
The reduced matrix elements $g^{q}_{[12]}$
and the reduced probability $B(q)\!\uparrow$ of excitation
induced by the external field (\ref{efpc})
are determined by the found values of $\rho^{q}_{(12)\eta}$
through the formulas:
\be
g^{q}_{[12]} =
\sum^{wind}_{[34]}
{\cal F}^{J_q}_{[12,34]}\,
\theta^{\vphantom{(+)}}_{(43)}\,
\delta^{\vphantom{(+)}}_{\eta_{\mbts{3}},\,-\eta_{\mbts{4}}}\,
\rho^{q}_{(34)\eta_{\mbts{3}}}\,,
\label{dbas6}
\ee
\be
B(q)\!\uparrow \,= \frac{1}{2J_q + 1}\,
\biggl|\,
\sum_{\eta} \sum^{wind}_{(12)}
\theta^{\vphantom{(+)}}_{(21)}\,
\bigl(\tilde{V}^0_{J_q} \bigr)_{(21)\eta}\,\rho^{q}_{(12)\eta}\,
\biggr|^2\,,
\label{dbas7}
\ee
where
\bea
\bigl(\tilde{V}^0_J \bigr)_{(21)\eta} &=& \sum_{LS}\,\eta^S
\bigl[\,u^{\vphantom{(2)}}_{(1)}\,v^{\vphantom{(2)}}_{(2)}
+ (-1)^S\,v^{\vphantom{(2)}}_{(1)}\,u^{\vphantom{(2)}}_{(2)}
\,\bigr]\,
\langle \,j_2 l_2\,||\,T_{JLS}\,||\,j_1 l_1\, \rangle \,
\bigl(\tilde{V}^0_{JLS} \bigr)_{(21)}\,,
\label{dbas8}\\
\bigl(\tilde{V}^0_{JLS} \bigr)_{(12)} &=&
\delta_{\tau_{\mbts{1}},\,\tau_{\mbts{2}}}
\int_0^{\infty} dr\,r^2\,R^{\vphantom{(2)}}_{(1)}(r)\,
R^{\vphantom{(2)}}_{(2)}(r)\,
\tilde{V}^0_{JLS\tau_{\mbts{1}}}(r)\,.
\label{dbas9}
\eea

\section{CORRELATED PROPAGATOR OF THE QTBA IN TERMS\\
         OF THE REDUCED MATRIX ELEMENTS \label{proprme}}

In this section we draw the formulas for the reduced matrix
elements of the correlated propagator
$A^{J\vphantom{(+)}}_{[12,34]} (\omega)$
which are obtained within the QTBA in case of the QPC model.
In what follows, the summation over single-quasiparticle indices
means summation over the discrete states
inside the ``pairing window''.
The general formula for the propagator of the QTBA
satisfying Eq.~(\ref{asprred}) reads:
\bea
A^{J\vphantom{(+)}}_{[12,34]} (\omega) &=&
\sum_{[5678]} \bigl[\,
\delta^{\vphantom{(+)}}_{[15]}\,\delta^{\vphantom{(+)}}_{[26]} +
Q^{J\,(+-)\,a}_{[12,56]} (\omega)\,\bigr]\,
A^{J\,(--)}_{[56,78]} (\omega)\,\bigl[\,
\delta^{\vphantom{(+)}}_{[73]}\,\delta^{\vphantom{(+)}}_{[84]} +
Q^{J\,(-+)\,a}_{[78,34]} (\omega)\,\bigr]
\nonumber\\
&&+\,{\txts \frac{1}{2}}\,
\bigl[\,P^{J\,(++)}_{[12,34]} (\omega) -
(-1)^{J+l_{\mbts{1}}-l_{\mbts{2}}+j_{\mbts{1}}-j_{\mbts{2}}}\,
P^{J\,(++)}_{[\bar{2}\bar{1},34]} (\omega)\,\bigr]\,,
\label{arm1}
\eea
where
\bea
Q^{J\,(+-)\,a}_{[12,34]} (\omega) &=&
\theta^{\vphantom{(+)}}_{(43)}
\bigl[\,Q^{J\,(+-)}_{[12,34]} (\omega) +
(-1)^{J+l_{\mbts{3}}-l_{\mbts{4}}+j_{\mbts{3}}-j_{\mbts{4}}}\,
Q^{J\,(+-)}_{[12,\bar{4}\bar{3}]} (\omega)\,\bigr]\,,
\label{qtlpm}\\
Q^{J\,(-+)\,a}_{[12,34]} (\omega) &=&
\theta^{\vphantom{(+)}}_{(21)}
\bigl[\,Q^{J\,(-+)}_{[12,34]} (\omega) +
(-1)^{J+l_{\mbts{1}}-l_{\mbts{2}}+j_{\mbts{1}}-j_{\mbts{2}}}\,
Q^{J\,(-+)}_{[\bar{2}\bar{1},34]} (\omega)\,\bigr]\,,
\label{qtlmp}
\eea
\be
A^{J\,(--)}_{[12,34]} (\omega) =
\delta^{\vphantom{(+)}}_{\eta_{\mbts{1}},-\eta_{\mbts{2}}}
\delta^{\vphantom{(+)}}_{\eta_{\mbts{3}},-\eta_{\mbts{4}}}
A^{J\,(--)}_{(12)\eta_{\mbts{1}},\,(34)\eta_{\mbts{3}}} (\omega)\,.
\label{ajmm}
\ee
The quantity $A^{J\,(--)}_{(12)\eta,\,(34)\eta'} (\omega)$
is a solution of the equation
\be
A^{J\,(--)}_{(12)\eta,\,(34)\eta'} (\omega) =
\tilde{A}^{J\vphantom{(-)}}_{(12)\eta,\,(34)\eta'} (\omega) +
\sum_{\eta''} \sum_{(56)}
\theta^{\vphantom{(+)}}_{(65)}\,
{\cal K}^{J}_{(12)\eta,\,(56)\eta''} (\omega)\,
A^{J\,(--)}_{(56)\eta'',\,(34)\eta'} (\omega)\,,
\label{arm2a}
\ee
where
\be
\tilde{A}^{J\vphantom{(-)}}_{(12)\eta,\,(34)\eta'} (\omega) =
- \frac{\eta\,\delta^{\vphantom{(+)}}_{\eta,\eta'}\bigl[
\delta^{\vphantom{(+)}}_{(13)}\,
\delta^{\vphantom{(+)}}_{(24)} +
(-1)^{J+l_{\mbts{1}}-l_{\mbts{2}}+j_{\mbts{1}}-j_{\mbts{2}}}\,
\delta^{\vphantom{(+)}}_{(14)}\,
\delta^{\vphantom{(+)}}_{(23)}\bigr]}
{2\,\bigl(\omega - \eta\,\bigl[E_{(1)} + E_{(2)}\bigr]\bigr)}\,,
\label{arm3a}
\ee
\vspace{1ex}
\be
{\cal K}^{J}_{(12)\eta,\,(34)\eta'} (\omega) =
\frac{\eta\,\bigl[
\Phi^{J\vphantom{(+)}}_{(12)\eta,\,(34)\eta'} (\omega) +
(-1)^{J+l_{\mbts{1}}-l_{\mbts{2}}+j_{\mbts{1}}-j_{\mbts{2}}}\,
\Phi^{J\vphantom{(+)}}_{(21)\eta,\,(34)\eta'} (\omega)\bigr]}
{\omega - \eta\,\bigl[E_{(1)} + E_{(2)}\bigr]}\,,
\label{kj}
\ee
\vspace{1ex}
\be
\Phi^{J\vphantom{(+)}}_{(12)\eta,\,(34)\eta'} (\omega) =
\sum_{\eta_{\mbts{1}}\eta_{\mbts{2}}\eta_{\mbts{3}}\eta_{\mbts{4}}}
\delta^{\vphantom{(+)}}_{\eta_{\mbts{1}},\,\eta\vphantom{\eta'}}\,
\delta^{\vphantom{(+)}}_{\eta_{\mbts{2}},-\eta\vphantom{\eta'}}\,
\delta^{\vphantom{(+)}}_{\eta_{\mbts{3}},\,\eta'}\,
\delta^{\vphantom{(+)}}_{\eta_{\mbts{4}},-\eta'}\,
\Phi^{J\vphantom{(+)}}_{[12,34]} (\omega)\,.
\label{phij}
\ee

It is easy to see that solution of the Eq.~(\ref{arm2a})
possesses the following symmetry:
\bea
A^{J\,(--)}_{(12)\eta,\,(34)\eta'} (\omega) &=&
(-1)^{J+l_{\mbts{1}}-l_{\mbts{2}}+j_{\mbts{1}}-j_{\mbts{2}}}\,
A^{J\,(--)}_{(21)\eta,\,(34)\eta'} (\omega)
\nonumber\\
&=&
(-1)^{J+l_{\mbts{3}}-l_{\mbts{4}}+j_{\mbts{3}}-j_{\mbts{4}}}\,
A^{J\,(--)}_{(12)\eta,\,(43)\eta'} (\omega)\,.
\label{ajmmpr}
\eea
It enables one to determine all the elements of the matrix
$A^{J\,(--)}_{(12)\eta,\,(34)\eta'} (\omega)$ by solving
Eq.~(\ref{arm2a}) for the non-zero block of the matrix
$\theta^{\vphantom{(+)}}_{(21)}\,
A^{J\,(--)}_{(12)\eta,\,(34)\eta'} (\omega)\,
\theta^{\vphantom{(+)}}_{(43)}$.
However, as follows from Eqs. (\ref{ajls3}), (\ref{adisc}),
(\ref{arm1})--(\ref{ajmm}), for the construction of
propagator in the coordinate representation
it is sufficient to determine the elements only of this block.

In order to define the remaining quantities which enter
Eqs. (\ref{arm1})--(\ref{qtlmp}), (\ref{phij})
in case of the quasiparticle-phonon coupling model
let us introduce notations:
\be
\Omega^{\vphantom{a}}_{12\,q} = E^{\vphantom{a}}_{12}
+ \eta^{\vphantom{(+)}}_1\,\omega^{\vphantom{(+)}}_q\,,\qquad
E^{\vphantom{a}}_{12} = \eta^{\vphantom{(+)}}_1\,
\bigl[\,E^{\vphantom{a}}_{(1)} +
E^{\vphantom{a}}_{(2)}\,\bigr]\,,
\label{esum}
\ee
\be
D^{\,q}_{[12,34]\,\eta} =
\delta^{\vphantom{q}}_{\eta,+1}\,
g^{q}_{[13]}\,g^{q^{\,\scs *}}_{[24]} +
\delta^{\vphantom{q}}_{\eta,-1}\,
(-1)^{j_{\mbts{1}}+j_{\mbts{2}}+j_{\mbts{3}}+j_{\mbts{4}}}\,
g^{q^{\,\scs *}}_{[31]}\,g^{q}_{[42]}\,,
\label{dfzq}
\ee
\be
X^{J\,q}_{[12,34]\,\eta} =
(-1)^{J+J_{\mbts{$q$}}+j_{\mbts{2}}-j_{\mbts{3}}}
\left\{
\begin{array}{ccc}
j_1 & j_2 & J \\ j_4 & j_3 & J_q \\
\end{array}
\right\}
D^{\,q}_{[12,34]\,\eta}\,,\quad
Y^q_{[12,3]} =
\frac{
\delta^{\vphantom{(+)}}_{j_{\mbts{1}}j_{\mbts{2}}}\,
\delta^{\vphantom{(+)}}_{l_{\mbts{1}}l_{\mbts{2}}}}
{2j_{\mbss{1}}+1}\,
D^{\,q}_{[12,33]\,\eta_{\mbts{3}}}\,,
\label{dfxq}
\ee
where the reduced matrix elements $g^{q}_{[12]}$ are
defined by the formulas of Sec.~\ref{qpi}.
Making use of the Eqs. (\ref{arme2}), (\ref{gmat6}),
(\ref{esum})--(\ref{dfxq}) we obtain:
\be
\Phi^{J\vphantom{\mbsu{(comp)}}}_{[12,34]} (\omega) =
\Phi^{J\,\mbsu{(res)}}_{[12,34]} (\omega) +
\Phi^{J\,\mbsu{(GSC)}}_{[12,34]} (\omega) +
\Phi^{\,\mbsu{(comp)}}_{[12,34]} (\omega)\,,
\vphantom{\sum_q}
\label{arm4}
\ee
\bea
\Phi^{J\,\mbsu{(res)}}_{[12,34]} (\omega) &=&
\eta^{\vphantom{(+)}}_1\,
\delta^{\vphantom{(+)}}_{\eta_{\mbts{1}},-\eta_{\mbts{2}}}\,
\delta^{\vphantom{(+)}}_{\eta_{\mbts{3}},-\eta_{\mbts{4}}}
\sum_q \biggl[\,
\delta^{\vphantom{(+)}}_{\eta_{\mbts{1}},\eta_{\mbts{3}}}\,
\biggl(
\frac{X^{J\,q}_{[12,34]\,\eta_{\mbts{1}}}}
{\omega-\Omega_{32\,q}} +
\frac{X^{J\,q}_{[12,34]\,\eta_{\mbts{2}}}}
{\omega-\Omega_{14\,q}}\biggr)
\nonumber\\
&&+\;
\delta^{\vphantom{(+)}}_{[24]} \sum_{[5]} \frac{
\delta^{\vphantom{(+)}}_{\eta_{\mbts{5}},\eta_{\mbts{1}}}\,
Y^q_{[13,5]}}{\omega-\Omega_{52\,q}} +
\delta^{\vphantom{(+)}}_{[13]} \sum_{[6]} \frac{
\delta^{\vphantom{(+)}}_{\eta_{\mbts{6}},\eta_{\mbts{2}}}\,
Y^q_{[42,6]}}{\omega-\Omega_{16\,q}}\,\biggl]\,,
\label{arm5}
\eea
\bea
\Phi^{J\,\mbsu{(GSC)}}_{[12,34]} (\omega) &=&
- \eta^{\vphantom{(+)}}_1\,
\delta^{\vphantom{(+)}}_{\eta_{\mbts{1}},-\eta_{\mbts{2}}}\,
\delta^{\vphantom{(+)}}_{\eta_{\mbts{3}},-\eta_{\mbts{4}}}
\sum_q \biggl[\,
\delta^{\vphantom{(+)}}_{\eta_{\mbts{1}},-\eta_{\mbts{3}}}\,
\biggl(\,
\frac{X^{J\,q}_{[12,34]\,\eta_{\mbts{1}}}}{\Omega_{42\,q}} +
\frac{X^{J\,q}_{[12,34]\,\eta_{\mbts{2}}}}{\Omega_{13\,q}}
\,\biggr)
\nonumber\\
&&+\;
\delta^{\vphantom{(+)}}_{[24]} \sum_{[5]} \frac{
\delta^{\vphantom{(+)}}_{\eta_{\mbts{5}},\eta_{\mbts{2}}}\,
Y^q_{[13,5]}\,(\omega - E_{12} - \Omega_{35\,q})}
{\Omega_{15\,q}\,\Omega_{35\,q}}
\nonumber\\
&&+\;
\delta^{\vphantom{(+)}}_{[13]} \sum_{[6]} \frac{
\delta^{\vphantom{(+)}}_{\eta_{\mbts{6}},\eta_{\mbts{1}}}\,
Y^q_{[42,6]}\,(\omega - E_{12} - \Omega_{64\,q})}
{\Omega_{62\,q}\,\Omega_{64\,q}}\,\biggl]\,,
\label{arm6}
\eea
\bea
\Phi^{\,\mbsu{(comp)}}_{[12,34]} (\omega) &=&
- \eta^{\vphantom{(+)}}_1\,
\delta^{\vphantom{(+)}}_{\eta_{\mbts{1}},-\eta_{\mbts{2}}}\,
\delta^{\vphantom{(+)}}_{\eta_{\mbts{3}},-\eta_{\mbts{4}}}\,
\delta^{\vphantom{(+)}}_{\eta_{\mbts{1}},\eta_{\mbts{3}}}
\sum_{[56]qq'}
\frac{
\delta^{\vphantom{(+)}}_{\eta_{\mbts{5}},\eta_{\mbts{2}}}\,
\delta^{\vphantom{(+)}}_{\eta_{\mbts{6}},\eta_{\mbts{1}}}\,
Y^q_{[13,5]}\,Y^{q'}_{[42,6]}}
{\Omega_{15\,q}\,\Omega_{35\,q}\,
\Omega_{62\,q'}\,\Omega_{64\,q'}}
\nonumber\\
&&\times
(\omega + E_{56} - \Omega_{12\,q} - \Omega_{34\,q'})\,,
\vphantom{\sum_{B_{C_D}}}
\label{arm7}
\eea
\bea
P^{J\,(++)}_{[12,34]} (\omega) &=&
\eta^{\vphantom{(+)}}_1\,
\delta^{\vphantom{(+)}}_{\eta_{\mbts{1}},\eta_{\mbts{2}}}\,
\delta^{\vphantom{(+)}}_{\eta_{\mbts{3}},\eta_{\mbts{4}}}
\sum_q \biggl[\,
\frac{
\delta^{\vphantom{(+)}}_{\eta_{\mbts{1}},-\eta_{\mbts{3}}}\,
X^{J\,q}_{[12,34]\,\eta_{\mbts{4}}}}
{\Omega_{13\,q}\,\Omega_{24\,q}}\,
\biggl(\frac{1}{\omega-\Omega_{32\,q}} -
\frac{1}{\omega-\Omega_{14\,q}}\biggr)
\nonumber\\
&+&
\delta^{\vphantom{(+)}}_{[24]} \sum_{[5]} \frac{
\delta^{\vphantom{(+)}}_{\eta_{\mbts{5}},-\eta_{\mbts{1}}}\,
Y^q_{[13,5]}}{\Omega_{51\,q}\,\Omega_{53\,q}\,
(\omega-\Omega_{52\,q})} -
\delta^{\vphantom{(+)}}_{[13]} \sum_{[6]} \frac{
\delta^{\vphantom{(+)}}_{\eta_{\mbts{6}},-\eta_{\mbts{2}}}\,
Y^q_{[42,6]}}{\Omega_{26\,q}\,\Omega_{46\,q}\,
(\omega-\Omega_{16\,q})}\,\biggl]\,,
\label{arm8}
\eea
\bea
Q^{J\,(+-)}_{[12,34]} (\omega) &=&
\delta^{\vphantom{(+)}}_{\eta_{\mbts{1}},\eta_{\mbts{2}}}\,
\delta^{\vphantom{(+)}}_{\eta_{\mbts{3}},-\eta_{\mbts{4}}}
\sum_q \biggl\{\,
\frac{
\delta^{\vphantom{(+)}}_{\eta_{\mbts{1}},-\eta_{\mbts{3}}}\,
X^{J\,q}_{[12,34]\,\eta_{\mbts{3}}}}{\Omega_{31\,q}\,
(\omega - \Omega_{32\,q})} +
\frac{
\delta^{\vphantom{(+)}}_{\eta_{\mbts{2}},-\eta_{\mbts{4}}}\,
X^{J\,q}_{[12,34]\,\eta_{\mbts{4}}}}{\Omega_{24\,q}\,
(\omega - \Omega_{14\,q})}
\nonumber\\
&+&
\delta^{\vphantom{(+)}}_{[24]} \sum_{[5]} \biggl[
\frac{
\delta^{\vphantom{(+)}}_{\eta_{\mbts{5}},\eta_{\mbts{1}}}\,
Y^q_{[13,5]}}{E_{13}\,\Omega_{53\,q}} +
\frac{
\delta^{\vphantom{(+)}}_{\eta_{\mbts{5}},-\eta_{\mbts{1}}}\,
Y^q_{[13,5]}}{\Omega_{51\,q}}\,\biggl(\,
\frac{1}{E_{13}} +
\frac{1}{\omega-\Omega_{52\,q}}\,\biggr)\,\biggr]
\nonumber\\
&+&
\delta^{\vphantom{(+)}}_{[13]} \sum_{[6]} \biggl[
\frac{
\delta^{\vphantom{(+)}}_{\eta_{\mbts{6}},\eta_{\mbts{2}}}\,
Y^q_{[42,6]}}{E_{42}\,\Omega_{46\,q}} +
\frac{
\delta^{\vphantom{(+)}}_{\eta_{\mbts{6}},-\eta_{\mbts{2}}}\,
Y^q_{[42,6]}}{\Omega_{26\,q}}\,\biggl(\,
\frac{1}{E_{42}} +
\frac{1}{\omega-\Omega_{16\,q}}\,\biggr)\,\biggr]\,\biggr\}\,,
\label{arm9}
\eea
\bea
Q^{J\,(-+)}_{[12,34]} (\omega) &=&
(-1)^{j_{\mbts{1}}+j_{\mbts{2}}+j_{\mbts{3}}+j_{\mbts{4}}}\,
Q^{J\,(+-)}_{[43,21]} (-\omega)
\nonumber\\
&=&
\delta^{\vphantom{(+)}}_{\eta_{\mbts{1}},-\eta_{\mbts{2}}}\,
\delta^{\vphantom{(+)}}_{\eta_{\mbts{3}},\eta_{\mbts{4}}}
\sum_q \biggl\{\,
\frac{
\delta^{\vphantom{(+)}}_{\eta_{\mbts{4}},-\eta_{\mbts{2}}}\,
X^{J\,q}_{[12,34]\,\eta_{\mbts{1}}}}{\Omega_{42\,q}\,
(\omega - \Omega_{32\,q})} +
\frac{
\delta^{\vphantom{(+)}}_{\eta_{\mbts{3}},-\eta_{\mbts{1}}}\,
X^{J\,q}_{[12,34]\,\eta_{\mbts{2}}}}{\Omega_{13\,q}\,
(\omega - \Omega_{14\,q})}
\nonumber\\
&+&
\delta^{\vphantom{(+)}}_{[24]} \sum_{[5]} \biggl[
\frac{
\delta^{\vphantom{(+)}}_{\eta_{\mbts{5}},\eta_{\mbts{3}}}\,
Y^q_{[13,5]}}{E_{31}\,\Omega_{51\,q}} +
\frac{
\delta^{\vphantom{(+)}}_{\eta_{\mbts{5}},-\eta_{\mbts{3}}}\,
Y^q_{[13,5]}}{\Omega_{53\,q}}\,\biggl(\,
\frac{1}{E_{31}} +
\frac{1}{\omega-\Omega_{52\,q}}\,\biggr)\,\biggr]
\nonumber\\
&+&
\delta^{\vphantom{(+)}}_{[13]} \sum_{[6]} \biggl[
\frac{
\delta^{\vphantom{(+)}}_{\eta_{\mbts{6}},\eta_{\mbts{4}}}\,
Y^q_{[42,6]}}{E_{24}\,\Omega_{26\,q}} +
\frac{
\delta^{\vphantom{(+)}}_{\eta_{\mbts{6}},-\eta_{\mbts{4}}}\,
Y^q_{[42,6]}}{\Omega_{46\,q}}\,\biggl(\,
\frac{1}{E_{24}} +
\frac{1}{\omega-\Omega_{36\,q}}\,\biggr)\,\biggr]\,\biggr\}\,.
\label{arm10}
\eea

The Eqs. (\ref{arm1})--(\ref{arm10}) completely define
the reduced matrix elements
$A^{J\vphantom{(+)}}_{[12,34]} (\omega)$
which enter Eqs. (\ref{ajls3}) and (\ref{adisc}) for the
``discrete'' part of the total correlated propagator in the
coordinate representation.
Notice that in the Eq.~(\ref{ajls3}) the matrix elements
$A^{J\vphantom{(+)}}_{[12,34]} (\omega)$
are the same for all the channels which differ
from each other only by the amplitudes $\alpha^c_{[12]}$.
Notice also that the form of the above equations for the
reduced matrix elements $A^{J\vphantom{(+)}}_{[12,34]} (\omega)$
are the same both for magic and for open-shell nuclei.
In the former case, however, the index
$\eta^{\vphantom{(+)}}_1$ in the set
$[1] = \{(1), \eta^{\vphantom{(+)}}_1 \}$
is not independent, but is determined by the occupation number
$n^{\vphantom{(+)}}_{(1)} =$ 0 or 1
of the state $(1) = (\lambda^{\vphantom{(+)}}_1)$ as:
$\eta^{\vphantom{(+)}}_1 = 1 - 2\,n^{\vphantom{(+)}}_{(1)}$.

\section{CALCULATIONS: DETAILS AND DISCUSSION \label{results}}

As an application of our approach isovector electric dipole
excitations in the GDR region have been calculated
in semi-magic isotopes $^{116,120,124}$Sn.
We started from a description of the independent single-particle
motion in the standard phenomenological Woods-Saxon potential.
Then the gap equation (\ref{gapeq}) was solved for neutron
subsystem under the usual condition that the number of particles
in valence zone is conserved on average.
In our calculations for the tin isotopes
this zone (``pairing window'') consists of all discrete
($\ve_{\lambda}<0$) and quasidiscrete states
above the chemical potential $\mu$ and of the discrete states
below $\mu$ starting from $1f_{7/2}$ subshell for neutrons
and $1d_{5/2}$ subshell for protons.
We emphasize that the same valence zone restricts summations
both in the gap equation (\ref{gapeq}) and
in the equations of Sec.~\ref{proprme} for the QTBA propagator.

As the quasidiscrete states we chose the discrete states with
$\ve_{\lambda}>0$ calculated with a box boundary condition
and having the last extremum of the radial wave function in the
range of non-vanishing values of the discrete spectrum
($\ve_{\lambda}<0$) wave functions.
According to this criterion the quasidiscrete states are selected
quite well if size of the box is not too large.
Notice that the partial components of the normal single-particle
GF $\tilde{G}^{\,nor}_{jl\tau} (r,r';\,\ve)$ entering
Eq.~(\ref{acont}) were calculated with outgoing wave boundary
condition except for the small number of components for which
$jl$-values coincide with $jl$-values of the quasidiscrete states.
For such partial components the box boundary condition was used
in order to avoid inconsistency in the calculations.

The parameter $c_p$ in Eq.~(\ref{parfxi}) was chosen so as to
obtain the averaged solution of the gap equation
$\bar{\Delta}_{\lambda}$ to be equal
to the averaged empirical value (see Ref.~\cite{BM1}):
\be
\bar{\Delta}=12\;\mbox{MeV} \times \mbox{A}^{-1/2}\,.
\ee
For $^{120}$Sn we have obtained
$\bar{\Delta} \approx 1.1\;\mbox{MeV}$ with $c_p$ = 0.719~MeV.
This parameter was used for the remaining two nuclei.
The quantity $\xi$ in Eq.~(\ref{parfxi}) is
determined as follows: $\xi = \sqrt{\xi_1\xi_2}$ where
$\,\xi_1 = \mu - \min(\ve^{\vphantom{(+)}}_{\lambda})$,
$\,\xi_2 = \max(\ve^{\vphantom{(+)}}_{\lambda}) - \mu$.

We assumed that the observable energies of single-particle
excitations in the neighbouring odd nuclei have to be reproduced
in the framework of the mean field plus BCS model.
These observable energies (experimental energy differences)
are defined by the following equation:
$
\ve^{exp}_{\lambda} = \pm\,
\bigl[\,{\cal E}^{\vphantom{'}}_{\lambda}
(\mbox{A} \pm 1) -
{\cal E}^{\vphantom{'}}_{GS}(\mbox{A})
\bigr]\,,
$
where ${\cal E}^{\vphantom{'}}_{GS}$ is the ground state
energy of the even-even nucleus (core) consisting of the
A nucleons, ${\cal E}^{\vphantom{'}}_{\lambda}$
is the energy of the ground or the excited state of the
neighbouring odd nucleus consisting of the
$\mbox{A} \pm 1$ nucleons.
In order to get an agreement with the experimental energies
the well depth of central part of the Woods-Saxon potential was
slightly varied so as to obtain
$\,\ve^{exp}_{\lambda}=\ve^{\vphantom{exp}}_{\lambda}$
for protons and
$\,\ve^{exp}_{\lambda}=\mu_{\tau_{\mbts{$\lambda$}}}\pm
E^{\vphantom{exp}}_{\lambda}$ for neutrons where
$\,E^{\vphantom{exp}}_{\lambda}$ is connected with
$\,\ve^{\vphantom{exp}}_{\lambda}$ by the Eq.~(\ref{dfqpen}).
Thus obtained energies and wave functions form the above mentioned
basis set $\{\ve_{\lambda}, \vphi_{\lambda}(x)\}$.

In the present calculations we included
ground state correlations (GSC) only in the QRPA part of the
correlated propagator. Another type of GSC caused by
quasiparticle-phonon coupling (GSC/QPC) and originated from
backward-going terms in time-ordered diagrams was not incorporated.
This means that:
(i) only the term $\Phi^{J\,\mbsu{(res)}}_{[12,34]} (\omega)$
in the Eq.~(\ref{arm4}) is accounted for;
(ii) associated components of the propagator containing
the functions
$\,P^{J\,(++)}_{[12,34]} (\omega)$,
$\,Q^{J\,(+-)}_{[12,34]} (\omega)$, and
$\,Q^{J\,(-+)}_{[12,34]} (\omega)$
[see Eqs. (\ref{arm1})--(\ref{qtlmp}), (\ref{arm8})--(\ref{arm10})]
are excluded from the calculations.

Since our single-particle and single-quasiparticle energies
are fitted to the experiment as it was mentioned above,
these energies and corresponding wave functions
already contain effectively the admixture of phonons.
This phonon contribution should be removed
from the mean field, energy gap,
and the effective interaction to avoid double counting
if the quasiparticle-phonon coupling is included explicitly.
To solve this problem we use the method which corresponds
to the self-consistent scheme of the QTBA (see Ref.~\cite{T05})
if the GSC/QPC are not included.
In this case the method consists of the replacement
of the amplitude $\Phi^{J\vphantom{(+)}}_{[12,34]} (\omega)$
in the Eq.~(\ref{phij}) by the difference amplitude
$\bar{\Phi}^{J\,\mbsu{(res)}}_{[12,34]} (\omega)$ where
\be
\bar{\Phi}^{J\,\mbsu{(res)}}_{[12,34]} (\omega) =
\Phi^{J\,\mbsu{(res)}}_{[12,34]} (\omega)
-\Phi^{J\,\mbsu{(res)}}_{[12,34]} (0)\,.
\label{barphi}
\ee
Making use of this subtraction procedure,
ground-state contributions of the
quasiparticle-phonon coupling defined by the quantity
$\Phi^{J\,\mbsu{(res)}}_{[12,34]} (0)$ are removed
from both the mass operator and the effective interaction.

In the calculations of the GDR
we neglected the contribution of pp and hh channels
that is justified for this type of excitations
(see Ref.~\cite{M67}). The parameters of the Landau-Migdal
zero-range interaction in the ph channel
entering Eqs. (\ref{intphst}) and (\ref{parfis})
were taken in accordance with the standard set
which is usually used in similar calculations
(see, e.g., Refs. \cite{KTT,KLLT}):
\be
f_{in} = -0.002\,,\qquad f'_{ex} = 2.62\,,\qquad f'_{in} = 0.76\,,
\nonumber
\ee
\be
g = 0.05\,, \qquad g' = 0.96\,, \qquad
C_0 = 300\;\,\mbox{MeV fm}^3\,.
\ee
Notice, however, that the parameter $f'_{ex}$
has been changed as compared to one used
in Refs. \cite{KTT,KLLT} in order to reproduce the
experimental mean energy of the GDR in $^{120}$Sn within
QTBA calculation. The parameter $f_{ex}$ was slightly varied
from $-2.187\,$ to $-1.957\,$
in the computation of the energies and amplitudes of
low-lying collective $\,2^+$ and $3^-$
phonons within QRPA in configuration space.
This allowed us to fit the energies of these phonons to
experimental values. Then the value of $f_{ex}$ was averaged
and used in the calculations of the
remaining phonons and further in the QTBA calculations.

The phonon characteristics were calculated within
the QRPA by making use of Eqs. (\ref{dbas1})--(\ref{dbas9}).
The configuration space was restricted by the
``pairing window'' defined above.
In this calculation the matrix elements of the
interaction ${\cal F}^{(c)S}_{(12,34)}$
with $S=1$ were omitted since they do not give
significant contribution in case of low-lying collective modes
with natural parity. In addition, we have neglected contribution
of pp and hh channels.
These simplifications decrease dimension of the QRPA matrix
in the configuration space by a factor of two.
For all three chosen tin isotopes
the collective modes with spin and parity
$\,2^+$, $3^-$, $4^+$, $5^-$, $6^+$,
and with energies below the neutron separation energy
were included into the phonon space.
A mode is assumed to enter the phonon space if its reduced
transition probability is more than 10~\% of the maximal one
for fixed spin and parity. This value was taken as an approximate
criterion for phonons selection. Characteristics of the low-lying
vibrations taken into account in the calculations are collected
in the Table~\ref{tab1}.
\begin{table}
\begin{center}
\caption{\label{tab1}
Energies  and reduced transition probabilities
of the low-lying phonons used in QTBA calculations.}
\vspace{3mm}
\tabcolsep=1.25em
\renewcommand{\arraystretch}{1.0}%
\begin{tabular}{ccccccc}
\hline
\hline
&\multicolumn{2}{c}{$^{116}$Sn \hphantom{abc}}
&\multicolumn{2}{c}{$^{120}$Sn \hphantom{abc}}
&\multicolumn{2}{c}{$^{124}$Sn \hphantom{abc}}\\
%\hline
J$^{\pi}$ & $\omega$  & B(EL)$\uparrow$ & $\omega$ & B(EL)$\uparrow$
& $\omega$ & B(EL)$\uparrow$\\
& (MeV) & (e$^2$fm$^{2L}) $ & (MeV) & (e$^2$fm$^{2L}) $ & (MeV)
& (e$^2$fm$^{2L}) $\\
\hline
2$^+$ & 1.29 & 2.97$\times$10$^3$ & 1.17 & 3.53$\times$10$^3$
& 1.13 & 2.55$\times$10$^3$\\
%\hline
3$^-$ & 2.27 & 9.06$\times$10$^4$ & 2.40 & 8.42$\times$10$^4$
& 2.60 & 1.32$\times$10$^5$\\
& 5.21 & 1.35$\times$10$^4$ & 5.07 & 1.89$\times$10$^4$ &&\\
%\hline
4$^+$ & 2.60 & 1.59$\times$10$^6$ & 2.62 & 1.22$\times$10$^6$
& 2.11 & 1.19$\times$10$^6$\\
& 6.56 & 5.15$\times$10$^5$ & 7.58 & 8.12$\times$10$^5$ & 3.93
& 5.87$\times$10$^5$\\
& 4.88 & 3.08$\times$10$^5$ & 5.35 & 3.35$\times$10$^5$ & 7.60
& 5.33$\times$10$^5$\\
& 5.40 & 3.02$\times$10$^5$ & 4.89 & 3.26$\times$10$^5$ & 5.06
& 4.17$\times$10$^5$\\
& 5.56 & 2.58$\times$10$^5$ & 3.44 & 2.25$\times$10$^5$ & 7.36
& 2.31$\times$10$^5$\\
&&&&& 5.38 & 1.64$\times$10$^5$\\
&&&&& 5.59 & 1.35$\times$10$^5$\\
&&&&& 4.48 & 1.30$\times$10$^5$\\
%\hline
5$^-$ & 5.91 & 2.95$\times$10$^7$ & 2.67 & 1.98$\times$10$^7$
& 6.85 & 2.85$\times$10$^7$\\
& 2.75 & 1.46$\times$10$^7$ & 7.04 & 1.80$\times$10$^7$ & 2.40
& 1.47$\times$10$^7$\\
& 6.13 & 6.53$\times$10$^6$ & 5.52 & 1.41$\times$10$^7$ & 5.49
& 1.41$\times$10$^7$\\
& 3.49 & 5.52$\times$10$^6$ & 6.53 & 9.98$\times$10$^6$ & 4.03
& 1.14$\times$10$^7$\\
& 8.39 & 4.15$\times$10$^6$ & 3.56 & 8.57$\times$10$^6$ & 8.48
& 9.66$\times$10$^6$\\
& 8.98 & 3.57$\times$10$^6$ & 8.59 & 8.43$\times$10$^6$ & 6.09
& 5.05$\times$10$^6$\\
& 3.82 & 3.13$\times$10$^6$ &&& 4.38 & 4.95$\times$10$^6$\\
%\hline
6$^+$ & 5.38 & 7.63$\times$10$^8$ & 7.44 & 9.74$\times$10$^8$
& 4.78 & 1.04$\times$10$^9$\\
& 5.70 & 7.51$\times$10$^8$ & 5.34 & 7.56$\times$10$^8$ & 7.24
& 8.79$\times$10$^8$\\
& 4.84 & 5.93$\times$10$^8$ & 4.27 & 5.63$\times$10$^8$ & 5.06
& 5.59$\times$10$^8$\\
& 6.41 & 5.48$\times$10$^8$ & 4.88 & 4.33$\times$10$^8$ & 2.40
& 5.47$\times$10$^8$\\
& 3.35 & 4.29$\times$10$^8$ & 2.94 & 3.93$\times$10$^8$ & 5.31
& 5.46$\times$10$^8$\\
& 3.99 & 2.27$\times$10$^8$ & 7.83 & 1.80$\times$10$^8$ & 7.67
& 2.58$\times$10$^8$\\
& 6.74 & 1.32$\times$10$^8$ & 4.98 & 1.67$\times$10$^8$ &&\\
& 3.70 & 1.00$\times$10$^8$ & 4.44 & 1.58$\times$10$^8$ &&\\
\hline
\hline
\end{tabular}
\end{center}
\end{table}

The dipole photoabsorption cross section $\sigma_{E1}(E)$
is the basic observable computed in the present work.
This value is expressed via the strength function $S_{E1}(E)$
according to the well known formula:
\be
\sigma_{E1}(E) = \frac{16 \pi^3 e^2}{9{\hbar}c} E S_{E1}(E)\,.
\label{sgth}
\ee
The strength function in turn is simply connected with the
polarizability $\Pi_{E1} (\omega)$ [see Eq.~(\ref{dfstrf})]:
\be
S_{E1}(E) = -\frac{1}{\pi} \mbox{Im}\,\Pi_{E1} (E + i\,\Delta)\,.
\ee
As follows from the Eq.~(\ref{dfpol4}),
to determine the value of function $\,\Pi_{E1} (\omega)$
at a given complex energy variable $\omega$
and, consequently, to compute the values of
$S_{E1}(E)$ and $\sigma_{E1}(E)$ one has to solve
Eq.~(\ref{lrmeqpc}) for the LRM. This equation was solved
within the framework of two models: QTBA and QRPA.
The spurious isoscalar $1^-$ state has been eliminated using
the ``forced consistency'' method presented in Sec.~\ref{sect4}.
The single-particle continuum was included as described
in Sec.~\ref{cpcr}.
Following calculational scheme which is usually used
in the response function formalism, in the present work
we introduced a smearing parameter $\Delta$ which is actually an
imaginary part of the energy variable $\,\omega$.
In the calculations of GDR the value $\Delta =$ 250 keV was used.

\begin{figure}[ht]
\includegraphics*[scale=1.6]{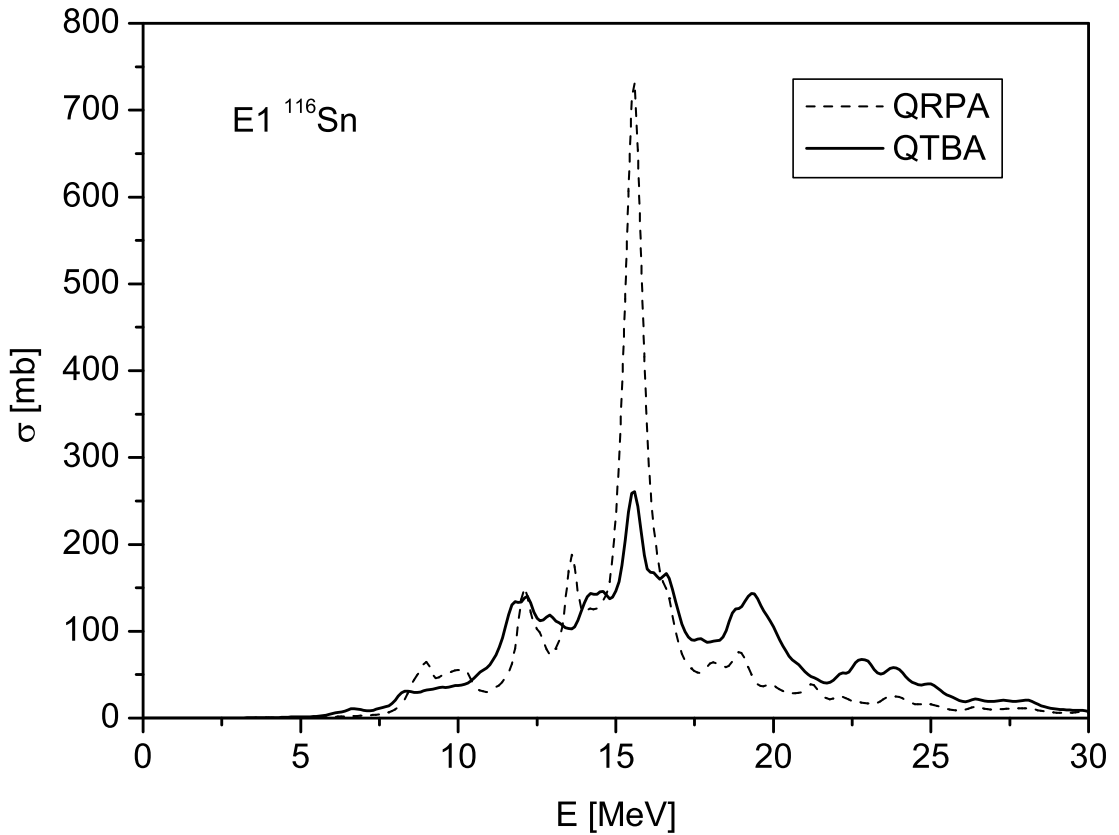}
\caption{\label{f1}
E1 photoabsorption cross section for $^{116}$Sn calculated
within QRPA (dashed line) and QTBA (solid line).
The smearing parameter $\Delta$ is equal to 250 keV.}
\end{figure}
\begin{figure}[ht]
\includegraphics*[scale=1.6]{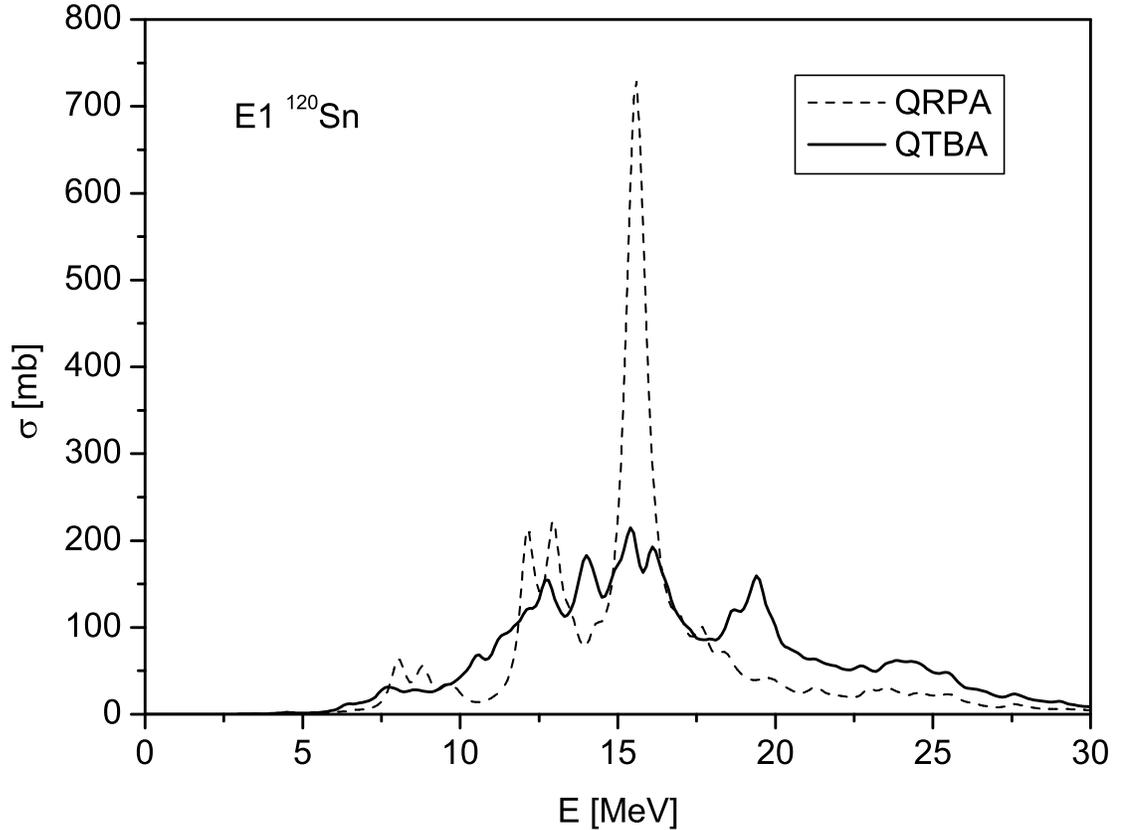}
\caption{\label{f2}
Same as Fig.~\ref{f1}, but for $^{120}$Sn.}
\end{figure}
\begin{figure}[ht]
\includegraphics*[scale=1.6]{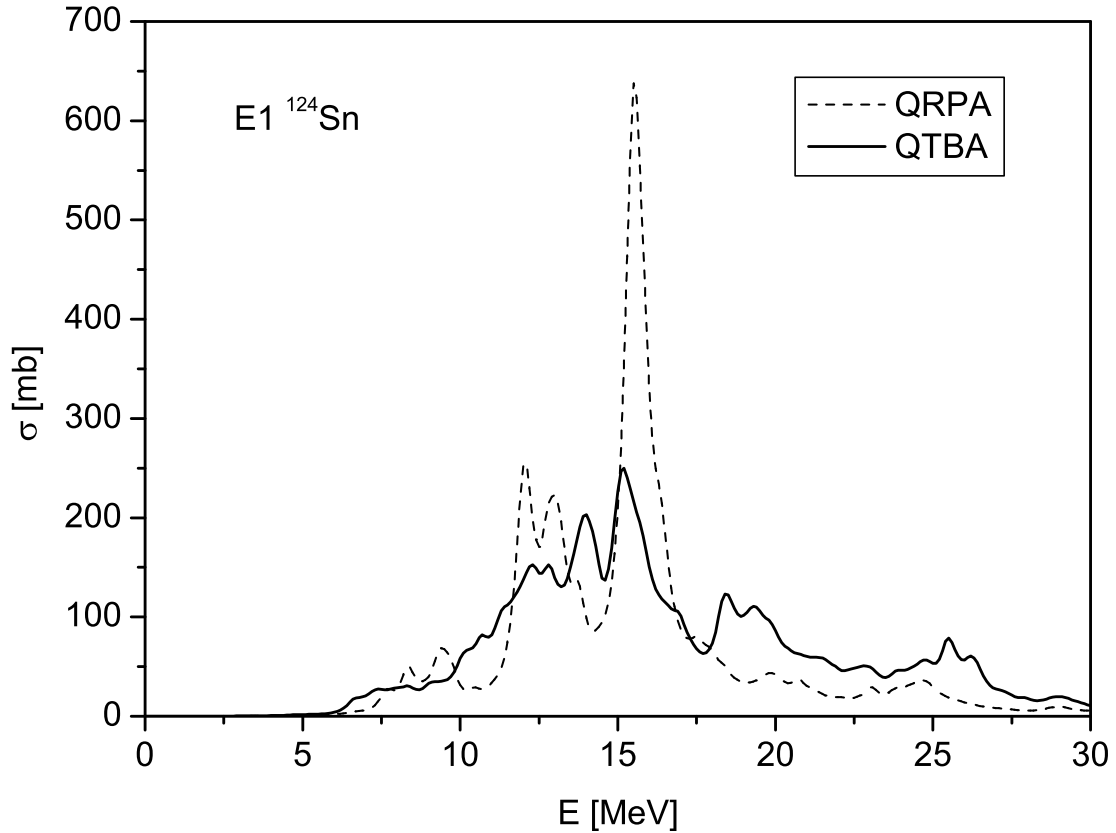}
\caption{\label{f3}
Same as Fig.~\ref{f1}, but for $^{124}$Sn.}
\end{figure}

Calculated photoabsorption cross sections
for three above indicated tin isotopes
are shown in Figs. \ref{f1}--\ref{f3}.
The solid and the dashed
curves represent the QTBA and the QRPA correspondingly.
In order to compare the  calculated cross sections with
experimental data the parameters of Lorentz function
$\sigma_L(E)$ were found where
\be
\sigma_L(E) = \sigma_0\,\frac{\Gamma^2 E^2}
{(E^2 - E_0^2)^2 + \Gamma^2 E^2}\,.
\label{ldis}
\ee
The parameters $E_0$, $\Gamma$, and $\sigma_0$
in the Eq.~(\ref{ldis})
were obtained by making use of the following condition:
the energy-weighted moments $m_0$, $m_{-1}$, and $m_{-2}$
of the functions (\ref{sgth}) and (\ref{ldis}) should coincide.
This method is analogous to the one developed in Ref.~\cite{TLor}
but in contrast to Ref.~\cite{TLor} we calculated the moments
in the finite energy interval 0--30 MeV
for which experimental data are available.
Parameters of the Lorentz fit are compiled in the Table~\ref{tab2}.
\begin{table}[ht]
\begin{center}
\caption{\label{tab2}
Lorentz function parameters of the GDR in
$^{116,120,124}$Sn obtained
within two microscopic approaches for 0--30 MeV energy interval.
Values of the depletion of the EWSR are presented as
percentages with respect to the corresponding TRK values.}
\vspace{3mm}
\tabcolsep=1.25em
\renewcommand{\arraystretch}{1.0}%
\begin{tabular}{lcccccc}
\hline
\hline
&\multicolumn{2}{c}{$^{116}$Sn}&\multicolumn{2}{c}{$^{120}$Sn}&
\multicolumn{2}{c}{$^{124}$Sn}\\
& QRPA & QTBA & QRPA & QTBA & QRPA & QTBA \\
\hline
$E_0$ (MeV) & 14.74 & 15.44 & 14.65 & 15.39 & 14.35 & 15.10 \\
%\hline
$\Gamma$ (MeV) & 2.4 & 4.0 & 2.7 & 4.4 & 2.6 & 4.4  \\
%\hline
$\sigma_0$ (mb) & 452 & 302 & 423 & 288 & 452 & 298  \\
%\hline
EWSR (\%) & 94 & 99 & 97 & 102 & 97 & 102  \\
\hline
\hline
\end{tabular}
\end{center}
\end{table}
In this table values of the depletion of the
energy weighted sum rule (EWSR), i.e. integrated cross sections
$\int_0^{\,30 \mbox{\scriptsize MeV}} \sigma_{E1}(E)\,dE$,
are also presented as percentages with respect to the
corresponding Thomas-Reiche-Kuhn (TRK) values 59.74 $NZ/A$ MeV mb.

In the Table~\ref{tab3} experimental Lorentz function parameters
of the GDR are shown.
\begin{table}[ht]
\begin{center}
\caption{\label{tab3}
Experimental Lorentz function parameters of the GDR.
The results are taken from
Refs. \protect\cite{74Le} and \protect\cite{69Fu}.}
\vspace{3mm}
\renewcommand{\arraystretch}{1.0}%
\begin{tabular}{lcccccc}
\hline
\hline
&\multicolumn{2}{c}{$^{116}$Sn}&\multicolumn{2}{c}{$^{120}$Sn}&
\multicolumn{2}{c}{$^{124}$Sn}\\
%\hline
& \protect\cite{74Le} & \protect\cite{69Fu}
& \protect\cite{74Le} & \protect\cite{69Fu}
& \protect\cite{74Le} & \protect\cite{69Fu}\\
\hline
$E_0$ (MeV) & 15.57 $ \pm$0.1 & 15.67$\pm$0.04
& 15.38$\pm$0.1 & 15.40$\pm$0.04
& 15.29$\pm$0.1 & 15.18$\pm$0.04\\
$\Gamma$ (MeV) & 5.21$\pm$0.1 & 4.19$\pm$0.06
& 5.25$\pm$0.1 & 4.88$\pm$0.06
& 4.96$\pm$0.1 & 4.81$\pm$0.06 \\
$\sigma_0$ (mb) & 270$\pm$5 & 266$\pm$7
& 284$\pm$5 & 280$\pm$8 & 275$\pm$5 & 283$\pm$8 \\
\hline
\hline
\end{tabular}
\end{center}
\end{table}
Experimental mean energies demonstrate the well known property
to decrease against neutron excess.
Our QTBA results drawn in the Table~\ref{tab2}
reproduce quite well these mean energies.
Since we take into account a finite number of the low-lying phonons,
our theoretical curves have the shapes which are rather
far from single Lorentzians. Nevertheless,
remaining parameters of the Lorentz function calculated
within QTBA, i.e. $\Gamma$ and $\sigma_0$,
are in a reasonable agreement with experimental values.

Consider the results obtained within QTBA and QRPA to analyze
effect of the QPC on the integral characteristics of resonances.
As one can see from the Table~\ref{tab2}, QTBA gives
significant increase of the total width as compared to QRPA
($\,\Gamma_{\mbss{QTBA}} \gtrsim 1.6\,\Gamma_{\mbss{QRPA}}$)
owing to contribution of the spreading width $\Gamma^{\downarrow}$.
Clearly this result could be expected from physical point of view.
The EWSR values obtained within the QTBA
for the investigated energy interval 0--30 MeV
are rather close to the TRK ones
but again there is a noticeable difference between
the QTBA and the QRPA values.
However this difference has another source which is
the subtraction procedure described above
[see Eq.~(\ref{barphi})]. It can be rigorously proved that for
the version of QTBA in which GSC/QPC are not taken into account
and the subtraction procedure is not applied the equality
EWSR$_{\mbss{QTBA}}$ = EWSR$_{\mbss{QRPA}}$ is fulfilled exactly.
Notice that the analogous equality is fulfilled between
the values of EWSR defined within the MCDD (Ref.~\cite{T89})
and within the RPA.
In our calculations just the subtraction procedure determined by
the Eq.~(\ref{barphi}) gives rise to increment of EWSR in the QTBA.
Switching off this procedure we have obtained that the values
of EWSR calculated within the QTBA in the ``infinite''
energy interval (0--200 MeV in our calculations)
are equal to the corresponding QRPA values with sufficiently
high accuracy. This result can be considered as a test of
our calculational scheme.
The mean energy shift of about 0.7--0.8 MeV obtained in QTBA
with respect to QRPA has the similar nature: it arises mainly
due to the subtraction procedure.
Without this subtraction the QTBA mean energies
decrease as compared to the QRPA ones by about of 0.2 MeV.
Thus, the subtraction procedure results in the significant
change of the averaged characteristics of the excited states
calculated in the QTBA as compared to the QRPA.
On the other hand, it ensures
elimination of the QPC contributions from the ground-state
characteristics leading to the equality between QTBA and QRPA
response functions $R^{\,\mbsu{eff}}(\omega)$ at $\omega = 0$.

Finally, notice that although we have taken
into account the most important effects of the QPC,
the neglected contributions of the GSC/QPC
and of the dynamical coupling to the pp and hh channels
may also affect the results.
In case of magic nuclei, the role of the GSC/QPC
in the description of nuclear excitations was investigated
in a series of papers (see, e.g., Ref.~\cite{KTT} and
references therein). The study of these effects in the nuclei
with pairing is in progress.

\section{CONCLUSION}

The quasiparticle time blocking approximation (QTBA)
is applied to describe E1 excitations in
the even-even open-shell spherical nuclei.
Within the QTBA pairing correlations,
two-quasiparticle (2q), and 2q$\otimes$phonon
configurations are included. The model is based on the
generalized Green function formalism in which the normal and
the anomalous Green functions are treated in a unified way
in terms of the components of generalized Green functions
in a doubled space.
To determine response of the spherically symmetric
system against an external field within the QTBA
the integral equation for the partial components of the
linear response matrix in the coordinate
representation has been obtained including coupling between
particle-hole, particle-particle, and hole-hole channels.
Configurations with a particle in the continuum are included
into the QRPA part of the response function.
This enables us to describe both spreading and escape widths
of nuclear excited states.
In our calculations we use phenomenological
Woods-Saxon single-particle input
and independently parametrized effective interaction
of the Landau-Migdal form.
So the additional procedure to eliminate spurious dipole
mode has been formulated in terms of QTBA.

The developed method has been applied to calculate isovector
E1 strength distribution in nuclei $^{116,120,124}$Sn.
The results for the photoabsorption cross sections in
the indicated tin isotopes are presented.
Since our main purpose was to test new approach,
these first calculations have been performed
assuming some additional simplifications of the model:
ground state correlations caused by quasiparticle-phonon coupling
and dynamical coupling to pp and hh channels were ignored.
Nevertheless noticeably fragmented giant dipole resonances
have been obtained for all three investigated nuclei.
Calculated integral characteristics of the resonances
are in a reasonable agreement with experimental values.

\vspace{1em}

\begin{flushleft}
{\normalsize\bf ACKNOWLEDGEMENTS}
\end{flushleft}
\vspace{0.5em}

E.~V.~L. acknowledges financial support from the
INTAS under the grant No. 03-54-6545 and
is grateful to A. von Humboldt Foundation
for financial support and to
Physik-Department der Technischen Universit\"at
M\"unchen for hospitality during completion of this work.
V.~I.~T. acknowledges financial support from the
Deutsche Forschungsgemeinschaft
under the grant No. 436 RUS 113/806/0-1
and from the Russian Foundation for Basic Research
under the grant No. 05-02-04005-DFG\_a.
E.~V.~L. is very thankful to S.~F.~Kovalev and G.~Ya.~Tertychny
for useful comments and discussion.
V.~I.~T. thanks the Institut f\"ur Kernphysik at the
Forschungszentrum J\"ulich for hospitality during the
completion of this work.

\end{document}